\documentclass[aps,floats,twocolumn,pre,10pt,groupedaddress,superscriptaddress]{revtex4-2}
\usepackage{amsmath,amssymb,amsthm,amsfonts,bm,bbm,braket,enumitem,mathtools}
\usepackage[english]{babel}
\usepackage{comment}
\usepackage{natbib}
\usepackage{hyperref}
\hypersetup{
    pdfauthor={Silvio Franz, Cosimo Lupo, Flavio Nicoletti, Giorgio Parisi and Federico Ricci-Tersenghi},
    pdftitle={Soft modes in vector spin glass models on sparse random graphs},
    colorlinks,
    linktocpage=true,
    pdfstartpage=1,
    pdfstartview=FitV,
    breaklinks=true,
    pageanchor=true,
    pdfpagemode=UseOutlines,
    plainpages=false,
    bookmarksnumbered,
    bookmarksopen=true,
    bookmarksopenlevel=1,
    hypertexnames=true,
    pdfhighlight=/O,
    urlcolor=blue,
    linkcolor=blue,
    citecolor=blue,
}

\allowdisplaybreaks

\usepackage[utf8x]{inputenc}	
\usepackage[T1]{fontenc} 	

\usepackage{graphicx}
\usepackage{array,booktabs}

\usepackage{soul}

\newcommand{\Hdat}{H_{\mathrm{dAT}}}
\newcommand{\Ctyp}{\mathcal{C}^{\mathrm{(typ)}}}

\begin{document}

\title{Soft modes in vector spin glass models on sparse random graphs}

\author{Silvio Franz}
\affiliation{LPTMS, UMR 8626, CNRS, Univ. Paris-Sud, Universit\'e Paris-Saclay, 91405 Orsay, France}

\author{Cosimo Lupo}
\email[Corresponding author: ]{cosimo.lupo89@gmail.com}
\affiliation{INFN, Sezione di Roma, 00185 Rome, Italy}

\author{Flavio Nicoletti}
\email[Corresponding author: ]{flavio.nicoletti@uniroma1.it}
\affiliation{Dipartimento di Fisica, Sapienza Universit\`a di Roma, 00185 Rome, Italy}

\author{Giorgio Parisi}
\affiliation{Dipartimento di Fisica, Sapienza Universit\`a di Roma, 00185 Rome, Italy}
\affiliation{INFN, Sezione di Roma, 00185 Rome, Italy}
\affiliation{CNR-Nanotec, Rome unit, 00185 Rome, Italy}

\author{Federico Ricci-Tersenghi}
\affiliation{Dipartimento di Fisica, Sapienza Universit\`a di Roma, 00185 Rome, Italy}
\affiliation{INFN, Sezione di Roma, 00185 Rome, Italy}
\affiliation{CNR-Nanotec, Rome unit, 00185 Rome, Italy}
\date{\today}

\begin{abstract}
We study numerically the Hessian of low-lying minima of vector spin glass models defined on random regular graphs. We consider the two-component (XY) and three-component (Heisenberg) spin glasses at zero temperature, subjected to the action of a randomly oriented external field. Varying the intensity of the external field, these models undergo a zero temperature phase transition from a paramagnet at high field to a spin glass at low field. We study how the spectral properties of the Hessian depend on the magnetic field. In particular, we study the shape of the spectrum at low frequency and the localization properties of low energy eigenvectors across the transition.  We find that in both phases the edge of the spectral density behaves as $\lambda^{3/2}$: such a behavior rules out the presence of a diverging spin-glass susceptibility $\chi_{SG}=\langle 1/\lambda^2 \rangle$. 
As to low energy eigenvectors, we find that the softest eigenmodes are always localized in both phases of the two models. However, by studying in detail the geometry of low energy eigenmodes across different energy scales close to the lower edge of the spectrum, we find a different behavior for the two models at the transition: in the XY case, low energy modes are typically localized; at variance, in the Heisenberg case low-energy eigenmodes with a multi-modal structure (sort of ``delocalization'') appear at an energy scale that vanishes in the infinite size limit. These geometrically non-trivial excitations, which we call Concentrated and Delocalised Low Energy Modes (CDLEM), coexist with trivially localised excitations: we interpret their existence as a sign of critical behavior related to the onset of the spin glass phase.
\end{abstract}

\maketitle

\section{Introduction}

The last years have seen a renewed interest in the properties of low-energy excitations of disordered solids and glasses. With respect to crystalline solids, low-temperature glasses are characterized by anomalous vibrational spectra displaying non-phononic low-frequency quasi-localized modes \cite{lerner2021low}. 
Phenomenological theories known as soft potential models \cite{gurevich2003anharmonicity, GurarieChalker2003, kuhn1997random, kuhn2003universality, kuhn2007finitely} predict a low-frequency non-phononic density of states (DoS) due to the presence of quasi-localized modes, which behave as $D(\omega)\sim A_4 \, \omega^4$ under plausible hypothesis.
Such a behaviour of the DoS is hidden in the phonon contribution, going as $\omega^2$ in three dimensions.  Numerical simulations of finite-dimensional models of structural glasses of small size \cite{lerner2016statistics, mizuno2017continuum, lerner2017effect, shimada2018anomalous, angelani2018probing, wang2019low, Wang2019sound, richard2020universality, bonfanti2020universal, ji2019, ji2020thermal, ji2021geometry} and random-field vector spin glasses \cite{baity2015soft, Thesis_Lupo2017} where phonons are absent, confirmed the presence of quasi-localized non-phononic modes with a quartic spectrum.
A mean-field version of the soft potential model \cite{rainone2021mean, bouchbinder2021low, folena2022marginal} with disordered coupling, also envisaged the same scenario. 
This picture however has been challenged for off-lattice systems, in a recent paper \cite{schirmacher2024nature} through an extension of heterogeneous elastic theory, which corroborated by numerical simulations, suggests that the low-frequency behavior of DoS is strongly sensitive to the details of the inter-particle interactions, the quartic law being an artifact of the tapering in the interaction at large distances.

Both the ubiquitous presence of the $\omega^4$ spectrum and the recent claim on its possible origin due to the microscopic details of the interaction potential, ask for first principle computations in solvable models, where the low-frequency spectrum can be computed analytically without resorting to effective models or approximations.
However, first principles theories and simple models, allowing us to understand the glassy soft modes beyond phenomenological assumptions, are just at the beginning. To this aim, spin glass models with quenched disorder and continuous degrees of freedom are a natural playground. At the mean-field level, spherical models, such as the $p$-spin \cite{Crisanti1992, cavagna1998stationary} and its related random-landscape model \cite{fyodorov2004complexity, fyodorov2018hessian} or the perceptron \cite{franz2015universal} are rather limited: the Hessian is very simple and belongs to the ensemble of rotational invariant random matrices, thus localisation is impossible. A richer description is provided by vector Sherrington-Kirkpatrick (SK) or vector $p$-spin glasses \cite{Book_MezardEtAl1987, Book_FischerHertz1991}. In this case the Hessian is a Rosenzweig-Porter random matrix \cite{porter1960statistical} that can exhibit localization at the edges of the spectrum \cite{lee2016extremal}. In these models, one finds two kinds of energy minima associated with stable and marginal glassy phases. Interestingly enough, the nature of low-energy excitations depends on the kind of glass one is dealing with \cite{franz2022delocalization, franz2022linear}. In stable glasses, the softest modes are localized on a single site, whereas they are delocalized in marginal glasses \cite{lee2016extremal, ikeda2023bose}. The spin-glass transition from stable to marginal glasses appears as a delocalization transition for low-energy modes \cite{franz2022delocalization}.

The most natural step to go beyond fully-connected models is to consider models defined on a Bethe lattice or random graph. These models are still solvable (because they have mean-field-like long-range interactions) but the finite connectivity (or graph degree) allows one to study the effect of spatial fluctuations and heterogeneities, which play a crucial role in the glass phenomenology. Unfortunately, while models with discrete Ising or Potts spins have been largely investigated \cite{klein1979, VianaBray1985, P.Mottishaw_1987, MezardParisi2001, MezardParisi2003, Krzakala_2008, Parisi_2014, Parisi2017, Concetti2018, Concetti2019, desantis2019computation, Perrupato2022}, only a little work has been done for vector spin glasses \cite{CoolenEtAl2005, LupoRicciTersenghi2017, LupoRicciTersenghi2018, LupoEtAl2019}.
In~\cite{Thesis_Lupo2017}, the low-temperature properties of the XY model were investigated, providing evidence for a quartic law in the spectrum of local energy minima, both in the paramagnetic and in the spin-glass phase. As a consequence, the zero-temperature spin-glass susceptibility $\chi_{SG}\equiv \int d\omega D(\omega)/\omega^4$ does not diverge at the transition in a field. This fact was already observed in \cite{sharma2016metastable} in the case of the Viana-Bray spin glass, and in \cite{baity2015soft} in the case of the three-dimensional Heisenberg spin glass. In \cite{folena2022marginal} a mechanism for a zero-temperature spin-glass transition with non-diverging susceptibility is described, in the context of the mean-field fully-connected soft potential model. In general, it appears that the zero-temperature transition in diluted models of vector spin glasses is different than that of fully-connected ones, since the spin-glass susceptibility of the latter is infinite in the whole spin-glass phase. In the absence of univocal indicators of criticality, the study of the properties of low-energy linear excitations could shed light on the mechanism underlying the transition.

In this work, we extend the study of vector spin glasses on a Random Regular Graph (RRG) in two directions. Firstly, we study the spectra of the Hessian at the energy minima of the Heisenberg ($m=3$ components) vector spin glass model. We find in this way that the validity of the quartic law is not accidental for the XY model, and support the hypothesis of it being a universal feature of disordered systems with finite connectivity interactions. The model has a zero-temperature spin-glass transition as a function of the field from a replica symmetric (RS) phase at high fields to a replica symmetry broken (RSB) phase at low fields. Secondly, we investigate in depth the change of nature of excitations in correspondence of the spin-glass transition, for both the XY and the Heisenberg models. We show that the properties of low-energy linear modes at criticality can be truly understood by studying their geometrical properties on the random graph, providing a novel method that goes beyond standard approaches based on the inverse participation ratio (IPR) and related quantities. At criticality, we find the presence of eigenmodes having a non-trivial multi-modes structure on the graph. Notably, eigenvectors exponentially localized on a sub-extensive number of distinct centers are found close to the edges of the spectrum of critical Erdos-Renyi~\footnote{Random graphs with Poisson-distributed degrees and average connectivity $c=O(\log N)$.} graphs \cite{Alt2021, tarzia2022fully}.

The structure of this paper is the following. In Section~\ref{sec:model_and_methods}, we briefly introduce the models and recap the approaches exploited for the study of their zero-temperature physics. Then, in Sections~\ref{sec:results_eigs} and \ref{sec:results_eigv}, we describe the low-frequency spectrum when varying the amplitude of the external field and we also study the topology of low-energy excitations, trying to understand if any localization/delocalization transition occurs in the graph in correspondence of the RS/RSB transition. Finally, in Section \ref{sec:concl} we draw our conclusions. Additional material is provided in the Appendices, in the Supplementary Information (SI) section.

\section{Model and methods}
\label{sec:model_and_methods}

We study the spin-glass model defined by the following Hamiltonian
\begin{equation}
\label{eq:Model_Hamiltonian}
    \mathcal{H}[\underline{{\mathbf{S}}}]\,=\,-\sum_{(i, j)\in\mathcal{E}}J_{ij}{\mathbf{S}}_i\cdot{\mathbf{S}}_j-H\sum_{i\in\mathcal{V}}{\mathbf{S}}_i\cdot{\mathbf{b}}_i
\end{equation}
where spins $\mathbf{S}_i$ are unit vectors with $m$ components, and $\mathcal{V}$ and $\mathcal{E}$ are respectively the vertex and edge sets of a random regular graph (RRG) with $N$ vertices and fixed connectivity $C=3$. The quenched couplings are independently chosen as $J_{ij}=\pm 1$ with equal probability, and the fields $\mathbf{b}_i$ are unit vectors independently drawn from the uniform distribution over the $m$-sphere.

The phase diagram of this model displays both a paramagnetic and a spin-glass phase, separated by the deAlmeida--Thouless (dAT) line in the $(H, T)$ plane, joining the $H=0$ critical point at a temperature $T_c$ and the zero-temperature transition point at a finite critical field $\Hdat$. In fully connected networks, the dAT line of vector spin glasses is measured analytically as explained in \cite{SharmaYoung2010}: only for spin glasses with $m>2$ components the critical field $\Hdat$ is finite. In sparse networks, the dAT line is finite for any $m\geq 1$ and is measured by studying the instability of the high-temperature (or high-field) Bethe solution of the model under perturbations \cite{ParisiEtAl2014, LupoRicciTersenghi2018, del2024compute}. 

The choice for the particular distributions of couplings and random fields we use 
is motivated by the fact that, at variance with fully-connected models, in sparse random graphs the spectral properties of the minima depend on the statistics of the amplitudes of the disorder parameters, $\{|J_{ij}|\}$ and $\{|\mathbf{b}_i|\}$. Our choice $|J|=1$ and $|\mathbf{b}|=1$ removes spatial heterogeneity of disorder strength, eliminating the most obvious disorder-related soft modes.

In the following, we study the $m=2$ (XY) and the $m=3$ (Heisenberg) spin glasses at zero temperature. We concentrate on low-lying energy minima, obtained through two different approaches for the two models.

\subsection{Minimisation algorithms}

\subsubsection{XY model}

In the case of the XY model, we use a combination of a message-passing (MP) or belief-propagation (BP) technique and a greedy minimization algorithm. 
We consider the zero-temperature replica-symmetric cavity equations for the ground state: these are recursive equations for the cavity fields $h_{i\rightarrow j}(\mathbf{S}_i)$ \cite{MezardParisi1987,MezardParisi2001}
\begin{multline}
    h_{i\rightarrow j}(\mathbf{S}_i)\,=\,-H \mathbf{S}_i\cdot\mathbf{b}_i \\
    -\sum_{k\in \partial i/j}\min_{\mathbf{S}_k}[J_{ik}\mathbf{S}_i\cdot\mathbf{S}_k-h_{k\rightarrow i}(\mathbf{S}_k)]
    \label{eq:BP_eqs_zeroT}
\end{multline}
where $\partial i$ is the set of neighbors of spin $i$.
We iterated these equations by discretizing the unit circle according to the prescription of the clock model \cite{NobreSherrington1986, NobreSherrington1989, IlkerBerker2013, IlkerBerker2014}, following the analysis in \cite{LupoRicciTersenghi2017, LupoRicciTersenghi2018, Thesis_Lupo2017}. When the iteration of these equations converges, the solution in the clock-model space can be used to identify the ground state for the original XY model with continuous varibles.
Convergence always takes place in the paramagnetic phase, where we obtain an estimate for the energy minimum which can be refined through a greedy over-relaxation algorithm (see the next paragraph). In the spin-glass phase, instead, the iterative approach does not converge; however, we still iterate the equation for a large number of sweeps and then we use the configuration thus obtained as the initial condition for the greedy algorithm. In both cases, we obtain much better minima (i.e.\ with lower energies) than by simply applying the greedy approach.

\subsubsection{Heisenberg model}

In the $m=3$ case, the use of the message-passing equations \eqref{eq:BP_eqs_zeroT} is much more involved, as it requires a smart discretization of the unit sphere. At variance with the unit circle, there is no unique uniform discretization of the unit sphere: for the sake of solving BP equations, a discretization that maximizes local uniformity is required, in order not to introduce biases in the iteration of the equations. This problem was tackled only very recently in \cite{del2024most} by some of us and it was not available when the present work was started.

We limit here to the use of the aforementioned greedy over-relaxation algorithm (GOA) \cite{baity2015soft, franz2022delocalization}. It consists of a sequence of local updates where field-alignment moves are combined with energy-preserving 180-degree inversions of the spins in the plane orthogonal to their molecular fields. In formulae, we update serially the spins according to the following rule
\begin{equation}
    \mathbf{S}_i \leftarrow \frac{\boldsymbol{\mu}_i+\mathcal{O}\,\mathbf{S}_i^{(R)}}{|\boldsymbol{\mu}_i+\mathcal{O}\,\mathbf{S}_i^{(R)}|}
    \label{eq:GD_with_OR}
\end{equation}
where
\begin{equation}
    \boldsymbol{\mu}_i = \sum_{j\in\partial i}J_{ij}\mathbf{S}_j+H\mathbf{b}_i
    \label{eq:local_field}
\end{equation}
is the molecular field acting on spin $i$, and
\begin{equation}
    \mathbf{S}_i^{(R)} \equiv \mathbf{S}_i^{\parallel}-\mathbf{S}_i^{\perp}
    \label{eq:S_refl}
\end{equation}
where $\mathbf{S}_i^{\parallel}$ and $\mathbf{S}_i^{\perp}$ are respectively the parallel and perpendicular components of $\mathbf{S}_i$ with respect to the molecular field $\boldsymbol{\mu}_i$, and $\mathcal{O}$ is the over-relaxation (OR) parameter. The $\mathcal{O}=0$ case corresponds to a purely
field-aligning greedy algorithm, while a non-zero value dumps the relaxation allowing to reach lower energy minima.
The GOA algorithm outperforms the simple greedy algorithm both in energy and sweep-convergence time, if the parameter $\mathcal{O}$ is properly set \cite{nicoletti2023low}.

\subsection{The Hessian}

The $N(m-1)\times N(m-1)$ Hessian matrix of small 
fluctuations of the spins around a minimum can be written as 
\begin{equation}
    \mathcal{M}_{ij}^{ab} = -(\mathbf{\hat{e}}_i^a \cdot \mathbf{\hat{e}}_j^b) J_{ij} + |\boldsymbol{\mu}_i|\delta_{ij}\delta_{ab} \, ,
    \label{eq:Hessian_reduced}
\end{equation}
where the vectors $\{\mathbf{\hat{e}}_i^a\}_{a=1}^{m-1}$ form  for each site $i$ an arbitrary orthonormal basis in the space orthogonal to spin~$\mathbf{S}_i$ \footnote{Note that in the case $m=2$ the orthogonal space is one-dimensional and thus \eqref{eq:Hessian_reduced} can be written only in terms of the angles $\{\theta_i^*\}$ identifying the spin configuration $\underline{\mathbf{S}}^*$ and the angles $\{\phi_i\}$ specifying the set of random external fields}.
To find numerically the low modes of the Hessians, we use the Arnoldi method~\cite{Arnoldi1951, Sorensen1992}, with a computational effort that for $M \times M$ sparse matrices like ours grows as $k M^2$, with $k$ being the number of eigenvectors to compute and $M=N (m-1)$. We restricted ourselves to the lowest $k=100$ eigenmodes of the Hessian for each sample.

\section{Numerical Results: eigenvalues}
\label{sec:results_eigs}

We measured low-energy harmonic spectra of XY and Heisenberg spin glasses, for several system sizes and field amplitudes.

For the simulations of the XY model, we have used four sizes, $N=10^n$ with $n \in \{3,4,5,6\}$, and fields $H=3.00, 2.50, 2.00, 1.50, 1.15, 1.10, 1.00, 0.60$. The critical field is $\Hdat=1.15(2)$, as estimated in \cite{LupoRicciTersenghi2017}. For each of these values of the field, we used $N_s=400, 200, 100, 50$ samples for the four sizes listed above, respectively. 

For the simulations of the Heisenberg model, we focused on five sizes, $N=2^n$ with $n \in \{12, 14, 16, 18, 20\}$, considering also sizes $N=2^{17}, 2^{19}$ for a few values of the external field. We considered field amplitudes $H=3.46, 2.60, 1.73, 1.30, 1.04, 0.87, 0.69, 0.52$. In this case, the critical field is $\Hdat=0.99(2)$ \cite{del2024compute}. For each of these values of the field, we used $N_s=5000, 2000, 500, 200, 50$ samples for the five sizes listed above, respectively. In addition to the measures of the spectral exponent, we studied the statistics of eigenvalues spacings in Appendix~\ref{sec:r}, finding results coherent with those presented in the remainder of the main text.

\subsection{The gapped region}

The spectrum of excitations has a gap if the magnetic field is strong enough.
It is easy to verify \footnote{As explained in Appendix~\ref{sec:gapped_phase} of the SI, the local fields actually have a non-zero lower bound for $H<c$, due to Onsager reaction.} that the local fields $\{\boldsymbol{\mu}_i\}$ \eqref{eq:local_field} of our model verify the bounds
\begin{equation}
    \max\{H-C,0\}\leq |\boldsymbol{\mu}_i| \leq H+C \, .
\end{equation}
For $H\gg C$, the lower edge of the spectrum behaves approximately as
\begin{equation}
    \lambda_{-}\simeq H-C-O(\sqrt{C})\equiv H-H_{\mathrm{gap}}
\end{equation}
as derived in Appendix~\ref{sec:gapped_phase} of the SI. Modes close to the lower spectral edge are strongly localized around vertices with local fields $|\mu_i-H+C|\ll 2C$.
At $H=H_{\mathrm{gap}}$, the spectral gap closes and soft modes appear. Since the focus of our work is on low-energy excitations, we leave to Appendix~\ref{sec:gapped_phase} the analysis of the gapped region.

\subsection{The lower tail of the spectrum}

The behavior of the spectral density $\rho(\lambda)$ in the vicinity of the lower edge $\lambda=0$ in the gapless regime yields information on the scaling of the DoS. Since $\lambda \equiv \omega^2$, 
\begin{equation}
    \rho(\lambda)\simeq K \lambda^a \quad \Leftrightarrow \quad \mathcal{D}(\omega)\simeq 2 K \omega^{2a+1} \, .
    \label{eq:rho_lambda_VS_DOS}
\end{equation}
To estimate the exponent $a$ and the prefactor $K$ from the data, we computed the sample-average of the smallest $100$ eigenvalues $\overline{\lambda_k}$, with $k$ being the rank of the mode, and estimated the cumulative distribution function $F(\overline{\lambda_k})$, which should behave as $F(\lambda)\simeq \frac{K}{a+1}\lambda^{a+1}$ at small~$\lambda$, by their relative rank $k/N (m-1)$.

\subsubsection{XY model}

In the left plot of the top panel of Fig.~\ref{fig:1} we show the cumulative distribution of Hessian eigenvalues in the minima of the XY spin glass, for $H$ ranging from $H \simeq 3 \, \Hdat$ to $H \simeq 0.9 \, \Hdat$, comprising both a region in the paramagnetic phase and a region in the spin-glass phase. We find that the lower tail exponent of the cumulative function is consistent with the value $1+a=5/2$ for all the values of the field, which corresponds to the $\omega^4$ spectrum. Consequently,  the zero-temperature spin-glass susceptibility $\chi_{SG}= \int \frac{\rho(\lambda)}{\lambda^2}d\lambda$ remains finite and regular at the transition. 

The prefactor $K$ depends on the field amplitude, as shown in the left figure of the bottom panel of Fig.~\ref{fig:1}, and has a non-monotonic dependence on $H$, attaining a maximum at $H \simeq 1.13 \, \Hdat$. Notice that $K$ ranges from about $0.02$ to about $0.10$ and remains of the same order of magnitude in the whole range of $H$. We find a qualitatively identical behavior when measuring the spectral prefactor for the minima obtained through applying solely the GOA minimization.

\subsubsection{Heisenberg model}

In the Heisenberg model, we cannot initialize GOA with the BP algorithm. Thus a careful choice of the over-relaxation parameter $\mathcal{O}$ is necessary to get good-quality low-energy minima. The higher the $\mathcal{O}$ value, the lower the energy of the minima reached by the GOA, but the longer the time to run the algorithm. In this model, we find that the best-fit exponent of the low-$\lambda$ spectrum depends on $\mathcal{O}$. However, for $\mathcal{O}$ sufficiently large the spectra cease to depend on $\mathcal{O}$ and are consistent with an exponent $a=3/2$. We find that $\mathcal{O}=10$ and $\mathcal{O}=50$ were large enough to achieve solid results, in the regimes of large fields ($H \gtrsim 1.5 \,\Hdat$) and close to the dAT line ($0.52 \,\Hdat \leq H \leq 1.30 \,\Hdat$) respectively.
Instead, lower values of $a$ are found for higher minima (reached using lower values of $\mathcal{O}$). We observe that less optimized minima correspond to a greater abundance of low-$\lambda$ modes. We concentrate hereafter on the lowest energy minima, discussing further this difference in Appendix~\ref{sec:robustness_spectral_exponent}. 

\begin{figure*}
    \centering
    \includegraphics[width=\textwidth]{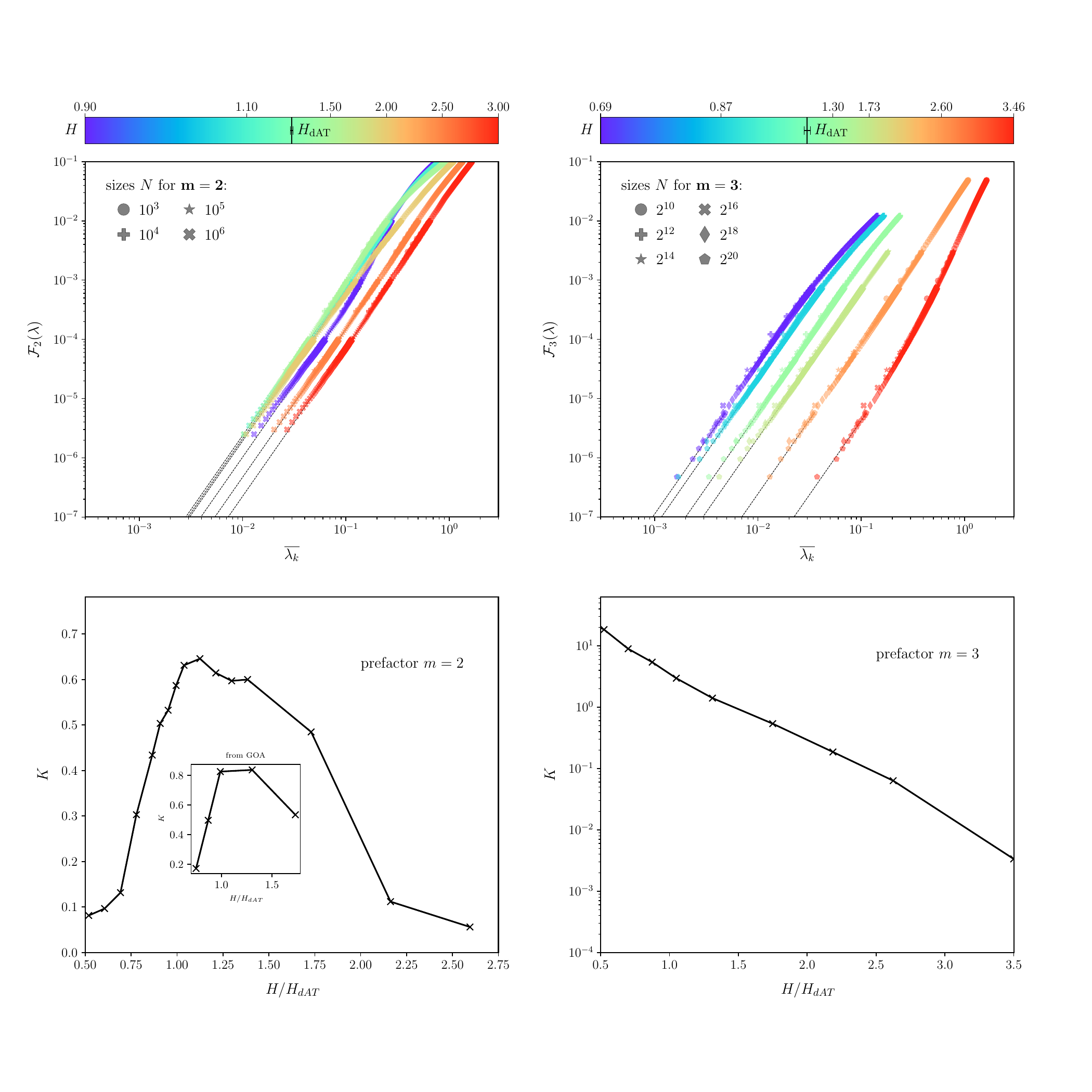}
    \caption{
        \textbf{Top}: measures of the cumulative distribution of eigenvalues: on the left, we have the XY model, on the right, the Heisenberg model. In both cases, the lower tail of the cumulative distribution is consistent with a $\lambda^{5/2}$ power-law behavior, corresponding to $\omega^4$ spectrum.
        \textbf{Bottom}: the prefactor $K$ as a function of $H/\Hdat$. This quantity has a very different behavior in the two models studied here: in the XY case, $K$ is a non-monotonic function of $H$, attaining a maximum close to the critical point; in the Heisenberg case, $K$ is a monotonic function and increases quite fast with decreasing $H$, spanning four orders of magnitudes for the range of external fields amplitude considered. The behavior observed in the XY model for the prefactor is qualitatively the same if minima are obtained through the GOA algorithm, as shown in the inset.
    }
    \label{fig:1}
\end{figure*}

In the right figure of the top panel of Fig.~\ref{fig:1} we show the cumulative distributions of eigenvalues for different values of $H$, ranging from $H \simeq 3 \,\Hdat$ down to $H \simeq 0.7 \,\Hdat$. The lower tail exponent is consistent with the value $1+a=5/2$, and also in this case the spin-glass susceptibility remains finite in the spin-glass phase.

As far as the spectral prefactor is concerned, we find here a very different behavior from the XY model. The prefactor is now a monotonic function of $H$ (see the bottom right panel of Fig.~\ref{fig:1}) spanning almost four orders of magnitude in the considered range $H \in [0.52 \Hdat,3.50\Hdat]$. As $H$ decreases, the Heisenberg spin glass acquires many more low-energy excitations than the XY one.

We notice that for both the XY and the Heisenberg models, the spectral density does not seem to contain a signal of the spin glass transition. This is at variance with fully connected vector spin glasses, where the spectral density exponent $a$ changes value at the transition~\cite{franz2022delocalization}.

\section{Numerical results: eigenvectors}
\label{sec:results_eigv}

We have seen that the shape of the spectrum has no qualitative changes at the transition. We would like now to investigate the behavior of the eigenmodes in the two phases, focusing in particular on the localization properties. 

Localization on a graph can have two aspects: \textit{i)} the fact that a large fraction of the eigenmode weight can be concentrated on a small number of sites, and if this holds, \textit{ii)} the sites over which the weight is concentrated can be all close in space. It is straightforward to notice that \textit{i)} does not necessarily imply \textit{ii)}. Indeed, due to the exponential proliferation of neighbors with the distance, tunneling between distant regions of the graph can occur if a coherent correlation between different paths is established.

The analysis of eigenvectors is divided into two parts: firstly, localization is studied by determining how the weight of the mode is distributed among the sites; then, we also take into account the underlying graph geometry, considering how the weight is distributed on the graph.

\subsection{Soft modes at the spin glass transition: non-geometrical properties}

We focus on localization properties of low-energy eigenmodes and we show in both models that no critical behavior at the transition is detectable from observables that ignore the underlying sparse structure of the graphs.

\subsubsection{The Inverse Participation Ratio}

A standard tool to characterize the degree of localization of a mode is the Inverse Participation Ratio (IPR), that in our model is defined as
\begin{equation}
    I(\lambda) \equiv \sum_{i=1}^N |\boldsymbol{\psi}_i(\lambda)|^4 \, ,
    \label{eq:IPR_def}
\end{equation}
where $\underline{\boldsymbol{\psi}}(\lambda)\,=\,(\boldsymbol{\psi}_1(\lambda),\dots, \boldsymbol{\psi}_N(\lambda))$ is the eigenvector related to eigenvalue $\lambda$ and each component is a $m-1$ dimensional vector $\boldsymbol{\psi}_i=(\psi_i^1,\dots,\psi_i^{m-1})$. Also notice the normalization choice for the eigenvectors, i.\,e. $\sum_i |\boldsymbol{\psi}_i(\lambda)|^2=1$. In the following, we will refer to the values of the square components $\{|\boldsymbol{\psi}_i|^2\}$ of an eigenvector as its \emph{weight vector} (WV).

The value assumed by the IPR in the infinite-size limit is equal to the inverse of the number of spins effectively participating in normalization and thus dominating the WV.
Hence, if the IPR is finite, then the related mode is dominated by a finite number of sites. This cluster of sites can be either localized in a finite region of the graph or scattered in multiple separated spots.
At variance, when the IPR goes to zero in the thermodynamic limit, the mode is always delocalized on the graph, since the mass is dominated by a cluster of diverging size.

\begin{figure}
    \centering
    \includegraphics[width=\columnwidth]{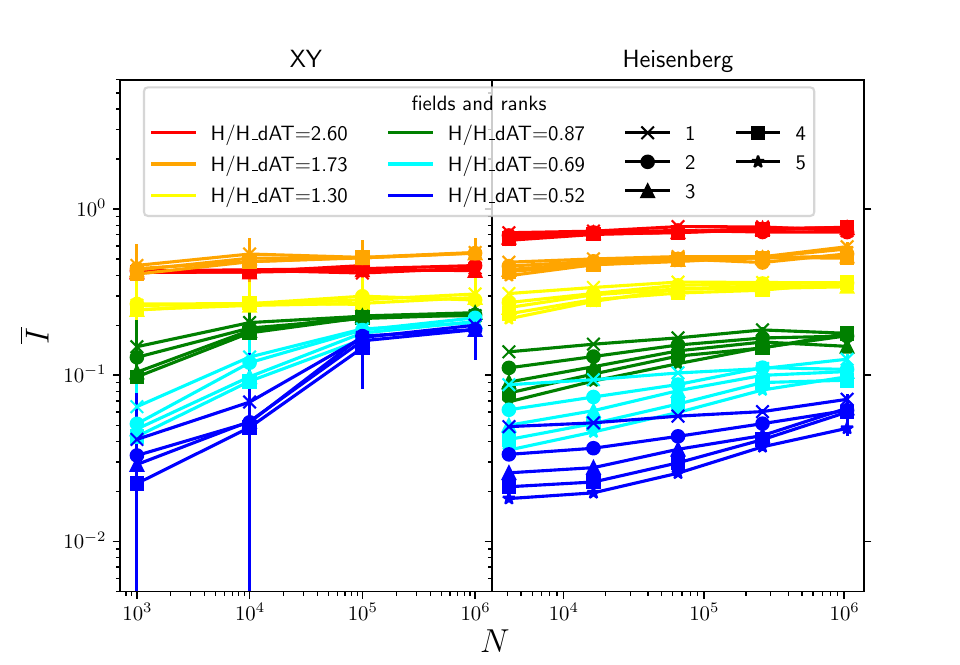}
    \caption{
        The average IPR of the softest modes, as a function of $N$: left the XY model, right the Heisenberg one. The IPR of low energy modes shows no criticality, as for both models in the large $N$ limit it remains finite at the transition. The large $N$ limit value of the IPR seems to increase with $H$.
    }
    \label{fig:average_ipr}
\end{figure}

We measured the sample average of the IPR of the rank-ordered soft modes for each size and field amplitude in our simulations. In Fig.~\ref{fig:average_ipr} we report our results: the average IPR of soft modes has a finite large-$N$ value both in the paramagnetic and in the spin glass phase, for both models. This result is opposite from what observed in the mean-field case for the Heisenberg model \cite{franz2022delocalization}, where the IPRs of soft modes vanish at the transition and in the whole spin-glass phase. The limiting values of the IPRs in Fig.~\ref{fig:average_ipr} get smaller for decreasing $H$, faster in the Heisenberg spin glass than in the XY one. Also, in the Heisenberg case, we observe that the properties of eigenmodes do not depend on the energy of the minimum reached by the OR. In Appendix~\ref{sec:robustness_eigenvectors}, we discuss the robustness of eigenvectors properties against GOA. 
Here and in the following, we show results regarding eigenvector properties related to $\mathcal{O}=10$ for $H/\Hdat \geq 1.73$ and $\mathcal{O}=50$ for $H/\Hdat \leq 1.30$.
 
\subsubsection{Multi-fractality of soft modes}

In order to have a complete picture of the non-geometrical properties of soft modes, we look at the probability distribution function (PDF) of the weights $|\boldsymbol{\psi}_i|^2$:
\begin{equation}
\label{eq:distr_weights}
    P(u)=\overline{\frac{1}{N}\sum_i \delta(u-|\boldsymbol{\psi}_i|^2)} \, .
\end{equation}
We found that $P(u)$ has a power law tail for large $u$ values, and thus we assume the following ansatz for the PDF
\begin{equation}
\label{eq:multi_fractal_0}
    P(u) \equiv N^{\gamma}\widetilde{P}(N^{\gamma} u) \, , \qquad \gamma>1
\end{equation}
with
\begin{equation}
\widetilde{P}(x)\simeq \frac{C_1}{x^{1+1/\gamma}} \qquad x \gg 1
    \label{eq:power-law-decay-eigvec-comps}
\end{equation}
The bound $u \le 1$ implies that the scaling variable $x$ is upper bounded by $N^\gamma$.

The moments of the PDF of the weights, also called generalized IPRs, are defined as
\begin{equation}
    I_q(\lambda) \equiv \sum_{i=1}^N |\boldsymbol{\psi}_i(\lambda)|^{2q} \, .
    \label{eq:generalised_IPRs}
\end{equation}
In the following, we show how the ansatz defined by eqs \eqref{eq:multi_fractal_0}, \eqref{eq:power-law-decay-eigvec-comps} affects the scaling of the moments in equation \eqref{eq:generalised_IPRs} with system size.
For $q>1$ the sum is dominated by the largest elements $|\psi_i|^2=O(1)$, or $x_i\sim N^\gamma$ and $I_q$ is finite and not scaling with $N$.
The moments with $q<1$ are instead more interesting and they can diverge in the limit $N\rightarrow\infty$.
When typical eigenvector square components have a scaling $O(1/N)$, it is immediate to verify from Eq.~\eqref{eq:generalised_IPRs} that $I_q\rightarrow \infty$ for any $q<1$.
When the typical component has a scaling $O(1/N^{\gamma})$ instead, with $\gamma>1$, then the moments in Eq.~\eqref{eq:generalised_IPRs} are divergent only for $q<1/\gamma\equiv q_*$.
The behavior of the moments $I_q$ can then be written as
\begin{equation}
    I_q(\lambda)\sim N^{(1-q/q_*)\theta(q_*-q)} \, .
    \label{eq:scaling_generalised_IPRs}
\end{equation}
This may resemble what is observed in multi-fractal localized modes of the Anderson model on small world lattices \cite{Garcia-Mata2020, GMata2022, rizzo2024localizedphaseandersonmodel}, although we show in the following that low energy excitations in the present vectorial models are different.

\begin{figure*}
    \centering
    \includegraphics[width=0.9\textwidth]{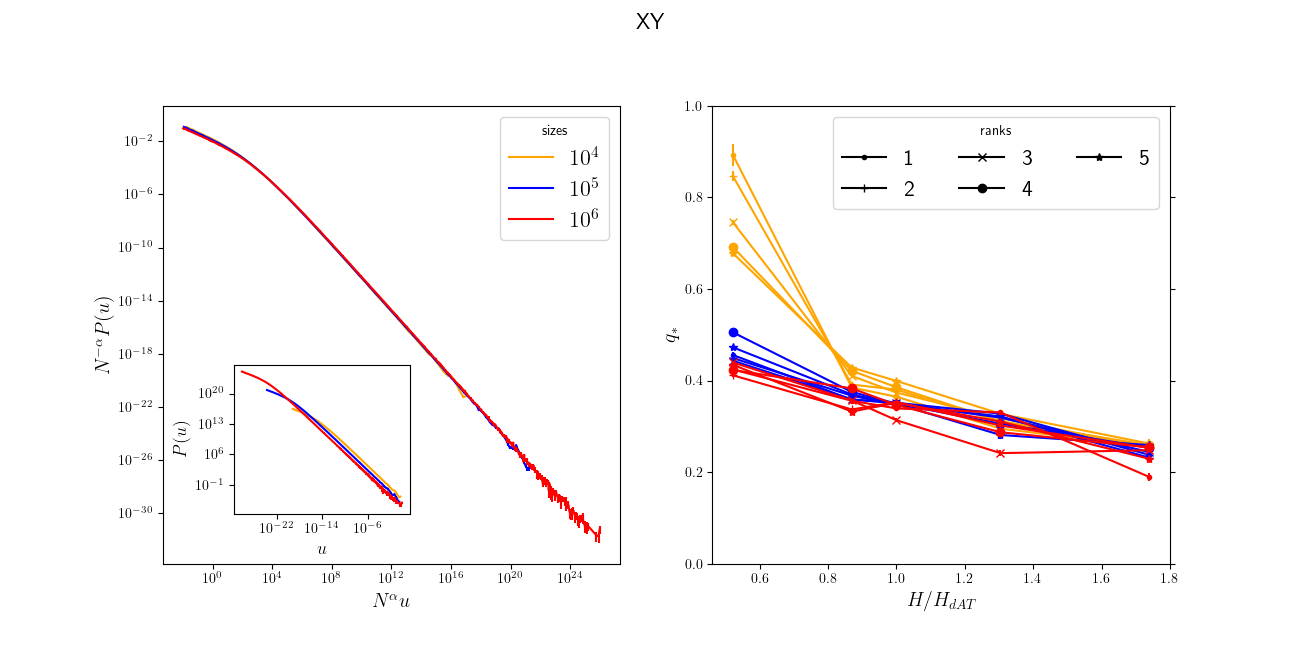}
    \caption{
        \textbf{Left}: the PDF of the rescaled square components $x=N^{\gamma} u$ of the smallest eigenmode, for sizes shown in legend and for $H=1.74 \Hdat$. The data shown are of the XY spin glass, as specified in the title. The empirical PDF is consistent with a power-law behavior with exponent $\beta\in[1,2]$, with $\beta \simeq 1.2$ for the value $H=1.74 \Hdat$. The inset shows the non-rescaled distribution $P_u(u)$.
        \newline
        \textbf{Right}: the exponent $q_*$ versus $H/\Hdat$. No critical behavior is observed for soft modes. The exponent $q_*$ of the smallest mode and size $N=10^6$ ranges from $q_*\simeq 0.2$ for $H=1.74 \Hdat$ to $q_*\simeq 0.42$ for $H=0.52 \Hdat$.
    }
    \label{fig:4XY}
\end{figure*}

\begin{figure*}
    \centering
    \includegraphics[width=0.9\textwidth]{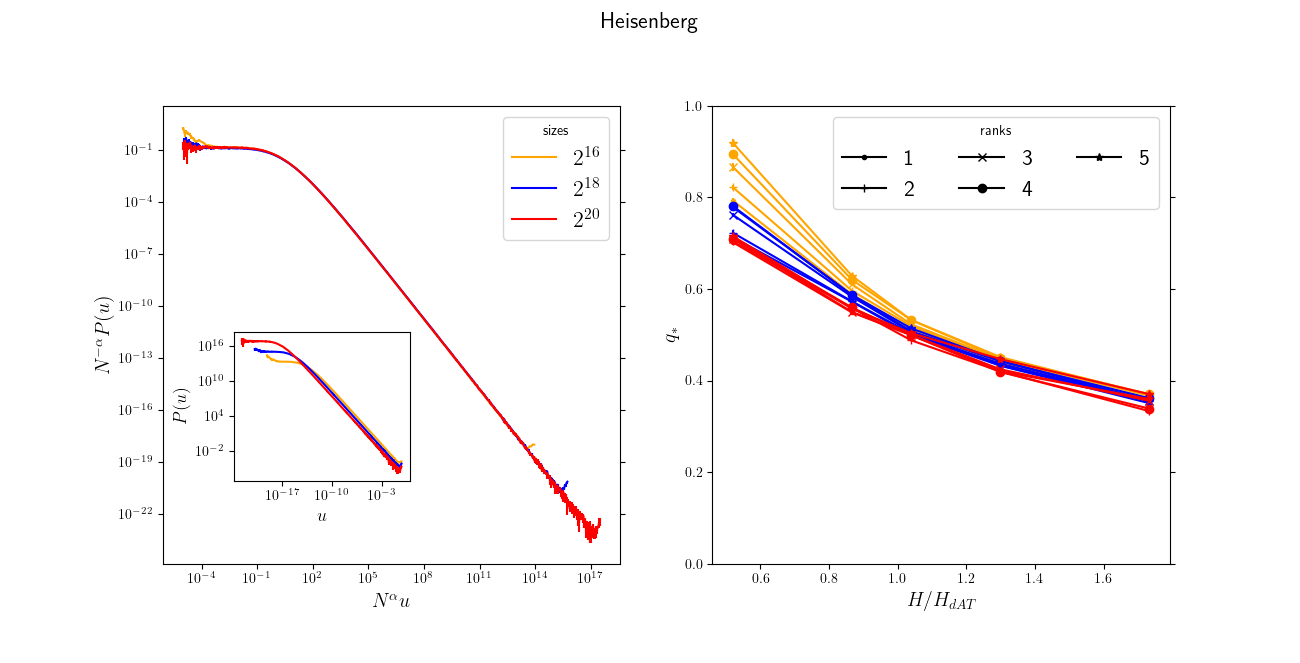}
    \caption{
        \textbf{Left}: the PDF of the rescaled square components $x=N^{\gamma} u$ of the smallest eigenmode, for sizes shown in legend and for $H=1.73 \Hdat$. The data shown are of the Heisenberg spin glass, as specified in the title. The empirical PDF is consistent with a power-law behavior with exponent $\beta\in[1,2]$, specifically $\beta\simeq 1.34$. The inset shows the non-rescaled distribution.
        \newline
        \textbf{Right}: the exponent $q_*$ versus $H/\Hdat$. No critical behavior is observed for the softest modes. The exponent $q_*$ of the smallest mode and size $N=2^{20}$ ranges from $q_*\simeq 0.34$ for $H=1.74 \Hdat$ to $q_*\simeq 0.67$ for $H=0.52 \Hdat$.
    }
    \label{fig:4Heis}
\end{figure*}

To fully characterize the multi-fractality of the soft modes of the XY and Heisenberg spin glass, for any size $N$, field $H$ and rank $k$ we measured the exponent of the tail of the PDF in Eq.~\eqref{eq:power-law-decay-eigvec-comps}. From the exponent, we can then obtain $q_*\equiv 1/\gamma$ as a function of $H$ for the softest modes.

We found similar results between the XY and the Heisenberg spin glasses. In Figs.~\ref{fig:4XY} and~\ref{fig:4Heis}, we show in the left panel of each figure the distribution $N^{-\gamma}P(N^{\gamma} u)$ for the lowest mode at $H = 1.73 \, \Hdat$, and for several sizes. The two insets report the corresponding unrescaled distribution $P(u)$. In both cases, the empirical PDFs are compatible with the ansatz in Eq.~\eqref{eq:power-law-decay-eigvec-comps}.

Moving the focus to higher ranks (right panels in Figs.~\ref{fig:4XY} and~\ref{fig:4Heis}), we found that $q_*$ depends quite weakly on~$k$ and~$N$ for $H>\Hdat$, for all ranks $k \in [1,100]$. This is not true for lower values of $H$: in the spin-glass phase, the dependence of $q_*$ on $k$ and $N$ is much more evident. However, even though $q_*$ varies sensibly with $k$ and $N$, we find no signature of criticality in the softest modes. Criticality would correspond to $q_*$ jumping or saturating to unity at $H = \Hdat$. Instead, the statistics of the softest modes seems to vary smoothly crossing the critical field. The emergence of long-range order at the transition does not correspond to a delocalization transition according to the usual criteria.

\subsection{Soft modes at the spin glass transition: geometrical properties}
\label{subsec:nongeom_sm}

The analysis of the localization properties of soft modes can be deepened by considering the spatial organization of the WV of each mode on the graph. We center the analysis on the site $i^*$ of maximum weight. We measure the total weight on the sites at a given distance $d$ from $i^*$~\footnote{The distance between two nodes is the minimal number of edges separating the two nodes.} uncovering that such a total weight does not decrease monotonically with the distance. Often a second local maximum in the weight vector exists and we study in detail the distance of such a second local maximum.

\subsubsection{Spatial distribution of the mass}

For a given eigenmode $\boldsymbol{\underline{\psi}}$, on a given graph, we define its weight spatial distribution (WSD) as follows
\begin{equation}
\label{eq:Correlation_around_main_core}
    \mathcal{C}(d)\,=\,\sum_{i\in s_d(i^*)}|\boldsymbol{\psi}_i|^{2} \, ,
\end{equation}
where $s_d(i^*)$ is the set of nodes at  distance $d$ from the node $i^*$ having the global maximum weight.

When the weight is localized close to the global maximum, the WSD decays rapidly with $d$. This is the generic situation that we find at high field amplitudes for soft modes. 
Typically, approaching a delocalization transition, the WSD decay on all distances would become slower and eventually one could find a $\mathcal{C}(d)$ growing with $d$ since the number of neighbors grows with the distance \cite{Garcia-Mata2020}.
In the present study, we never find such a delocalization pathway.
Instead, we observe that the WSD always decays at short distances but then it achieves a second maximum at large distances $d\sim \log(N)$. Such a WSD describes a soft mode having more than a single local maximum: although the weight is concentrated around each different local maxima, these weight local peaks are very far away from each other.
These very peculiar geometric features suggest the name Concentrated and Delocalized Low Energy Modes (CDLEM).
We claim that the existence of such CDLEM in the spin-glass phase constitutes a signal of long-range order, therefore we investigate their presence in the two models at study.

We measured the WSD of each eigenvector of both models. We then computed, for fixed $k, N, H$, the typical WSD
\begin{equation}
\label{eq:typ_WSD}
    \Ctyp(d) \equiv \exp \overline{\log \mathcal{C}(d)} \, .
\end{equation}
We looked for the long-distance peak of $\Ctyp(d)$ and found that, when present, this always occurs at a distance $L$ close to the graph diameter and more precisely at the distance $d_M$ where the graph has the maximum number of pairs of vertices
\begin{equation}
\label{eq:asymptotic_diameter_3RRG}
    L \approx d_M=O\left(\frac{\log(N)}{\log(c-1)}\right) \, .
\end{equation}
The exact expression of $d_M$ in terms of $N$ and $c$ can be found in Appendix~\ref{sec:properties_random_regular}.
In the large $N$ limit, it corresponds to the distance between two randomly chosen nodes~\cite{tishby2022mean}.

\begin{figure*}
\centering
    \includegraphics[width=0.49\textwidth]{figure_NEW/Corr0_AND_C2typ_XY.pdf}
    \includegraphics[width=0.49\textwidth]{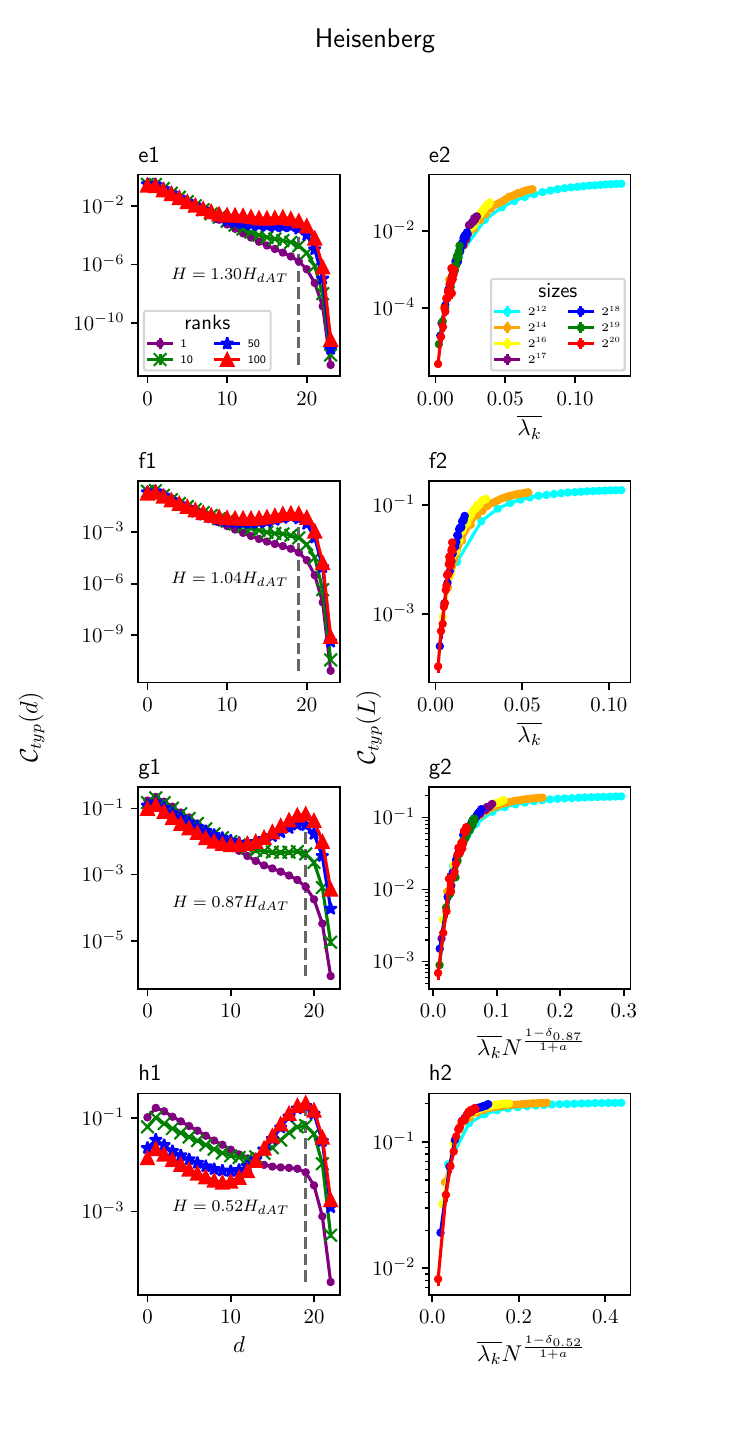}
    \caption{\textbf{Left}: Measures of the typical WSD for the XY model: figures a1, b1, c1, d1 represent this quantity as a function of distance on the graph, for system size $N=10^6$ and four different values of ranks, as indicated in the legend; a dashed vertical line indicates the position expected for the long-distance maximum at $L$. Figures a2, b2, c2, d2 show the long-distance peak of the WSD as a function of the average energy of the mode. The colors of the curve refer to different system sizes, as indicated in the legend.
    \textbf{Right}:  Measures of the typical WSD for the Heisenberg system: figures e1, f1, g1, h1 represent this quantity as a function of distance on the graph, for system size $N=10^6$ and four different values of ranks, as indicated in the legend; a dashed vertical line indicates the position expected for the long-distance maximum at $L$. Figures e2, f2, g2, h2 show the long-distance peak of the WSD as a function of the average energy of the mode. The colors of the curve refer to different system sizes, as indicated in the legend.
    }
    \label{fig:5}
\end{figure*}

The value assumed by the typical WSD \eqref{eq:typ_WSD} at $d=L$ is a proxy to monitor how much of the WV is concentrated around the maximum in $i^*$. For fully extended modes in the bulk of the spectrum, for which the WV is roughly flat, it is easy to see using Eq.~\eqref{eq:Correlation_around_main_core} that $\Ctyp(L)\simeq \mathcal{N}(L)/N\simeq 0.249$, where $\mathcal{N}(d)$ is the number of nodes at distance $d$ (more details in Appendix~\ref{sec:properties_random_regular}). For spatially localised modes, instead, $\Ctyp(L) \ll \mathcal{N}(L)/N$. Therefore, we measured $\Ctyp(L)$ as a function of average lower edge eigenvalues $\overline{\lambda_k}\simeq\left(\frac{(1+a)k}{K (m-1) N}\right)^{1/(1+a)}$, with $a=3/2$ and the prefactor $K$ measured in Sec.~\ref{sec:results_eigs}. This allows us to understand how the geometrical localization properties of typical modes change in different regions of the spectrum of low-energy excitations.

In Fig.~\ref{fig:5} we show a summary of our results for several external fields:
$H/\Hdat=1.30, 1.00, 0.87, 0.52$ for XY, and $H/\Hdat=1.30, 1.04, 0.87, 0.52$ for Heisenberg.
We plot $\Ctyp(d)$ versus $d$ in subfigures a1, b1, c1, d1 (XY) and e1, f1, g1, h1 (Heisenberg) for the largest size considered in our simulations, $N=10^6$ (XY) and $N=2^{20}$ (Heisenberg) respectively, and we show curves related to ranks $k=1, 10, 50, 100$. 
Moreover, we plot $\Ctyp(L)$ versus $\overline{\lambda_k}$ in subfigures a2, b2, c2, d2 (XY) and e2, f2, g2, h2 (Heisenberg), considering all sizes of our simulations.
Let us comment the results for the two models separately.

\paragraph*{XY.}
The typical WSD $\Ctyp(d)$ for $N=10^6$ does not show any local maximum at $d=L$ (always indicated with the vertical dashed line) at the transition field $\Hdat$, for any mode in the range $k\in[1, 100]$. One has to go deep into the spin-glass phase, e.g.\ $H=0.52\,\Hdat$, to observe the emergence of a local maximum at $d=L$, for modes with rank $k \geq 50$.
The impression is that, although some weight accumulates at $d=L$, this is not enough to produce a delocalization with a finite fraction of the weight moving away from the global maximum.

To quantify the above intuition, we observe that $\Ctyp(L)$ smoothly depends on $\overline{\lambda_k}$ in both phases and for all the sizes we have studied.
Assuming this scaling persists in the large $N$ limit, we can conclude that at the lower band edge ($\lambda_k \rightarrow 0$) the weight $\Ctyp(L)$ goes to zero and the corresponding eigenmodes are localized around the global maximum.
To obtain a finite value of $\Ctyp(L)$, and a secondary local maximum with a non-vanishing weight, one has to consider $k=O(N)$, i.e.\ eigenvectors in the bulk of the spectrum.

We can conclude that soft modes of the XY model are typically localized around a single site even in the glassy phase, with no evidence of criticality and delocalization at the transition.

\paragraph*{Heisenberg.}
In this case, although the typical WSD of the softest mode ($k=1$) still does not develop a local maximum at $d=L$ for any value of the field amplitude, this new local maximum appears for higher ranks as clearly shown in panels from e1 to h1 of Fig.~\ref{fig:5}.
This is a very strong hint that delocalization is taking place.
However, the key question becomes now how large the range must be to have delocalization (remember that in the XY model, the rank must be extensive and this means leaving the edge and entering the bulk of the spectrum).
We are going to show that for $1 \ll k \ll N$, i.e.\ $k = O(N^{\delta})$ with $\delta < 1$, delocalization takes place as the critical field is approached.

To quantify the scaling of the rank that leads to delocalization we study again $\Ctyp(L)$ as a function of $\lambda_k$. In the paramagnetic phase, the data measured in systems of different sizes collapse without the need for scaling (exactly as in the XY model), see e.g.\ panel e2 in Fig.~\ref{fig:5}.
In the spin-glass phase, instead, the data from different sizes collapse only if plotted as a function of
\begin{equation}
    \overline{\lambda_k} \, N^{(1-\delta)/(1+a)} \, , \quad \text{with } 0<\delta<1 \, .
    \label{eq:critical_scale_lambda}
\end{equation}
The case $\delta=1$ would correspond to extensive ranks $k=O(N)$ and data collapse without scaling $\overline{\lambda_k}$ (as in the XY model and the paramagnetic phase).
In the spin-glass phase, $\delta<1$ is required to properly collapse the data and this implies delocalization takes place in soft modes with $k=O(N^\delta)$.
These modes concentrate on the lower band edge in the large $N$ limit, that is they have arbitrarily low energies in the spectrum.

The values of the exponent $\delta$ get smaller lowering $H$: we have $\delta=1$ for $H\ge \Hdat$, $\delta=0.65$ for $H=0.87 \Hdat$ and $\delta=0.45$ for $H=0.52 \Hdat$. 

Looking at typical modes for given values of $k, N, H$, we conclude that CDLEMs do exist in the spin-glass phase. Such low-energy excitations appear to be rare in the XY model while are typical in the Heisenberg model, for sufficiently large ranks $k$. We will quantify in the next Section the probability of the appearance of CDLEM in the spin glass phase depending on the system size and energy distance from the lowest soft mode.

\begin{figure}
    \includegraphics[width=0.65\columnwidth]{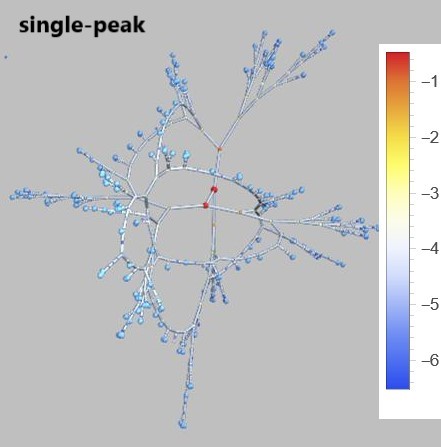}
    \includegraphics[width=0.65\columnwidth]{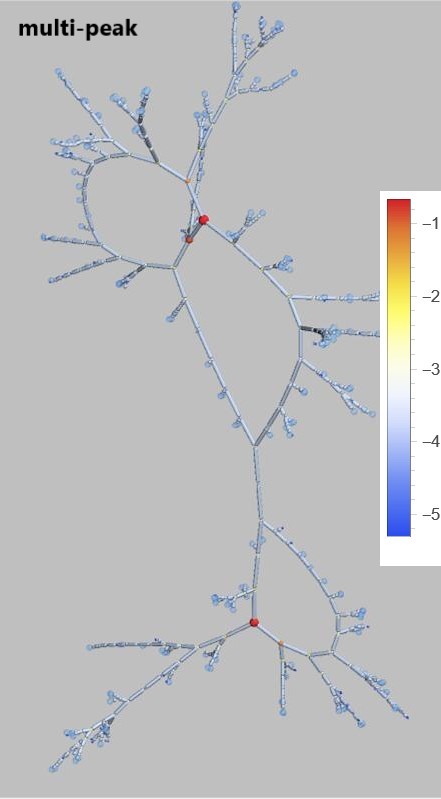}
    \caption{
        Visual representation of low energy modes with a single peak in the WSD (top panel) and two peaks in the WSD (bottom panel).
        \textbf{Top}: A single-peak low energy mode of the $m=3$ system, for $N=2^{14}$ and $H=1.73$. For this excitation, we have an IPR $I_2\simeq 0.17 \gg 1/2^{14}\simeq 0.00006$. We show the giant component of the subgraph constituted by the square components $u\geq 10^{-6}$.
        \textbf{Bottom}: A multi-peak low energy mode of the $m=3$ system, for $N=2^{14}$ and $H=1.73$. Here $I_2\simeq 0.09 \gg 1/2^{14}\simeq 0.00006$. We show the giant component of the subgraph constituted by the square components $u\geq 10^{-5}$.
    }
    \label{fig:visualized_modes}
\end{figure}

\begin{figure*}
    \centering
    \includegraphics[width=0.75\textwidth]{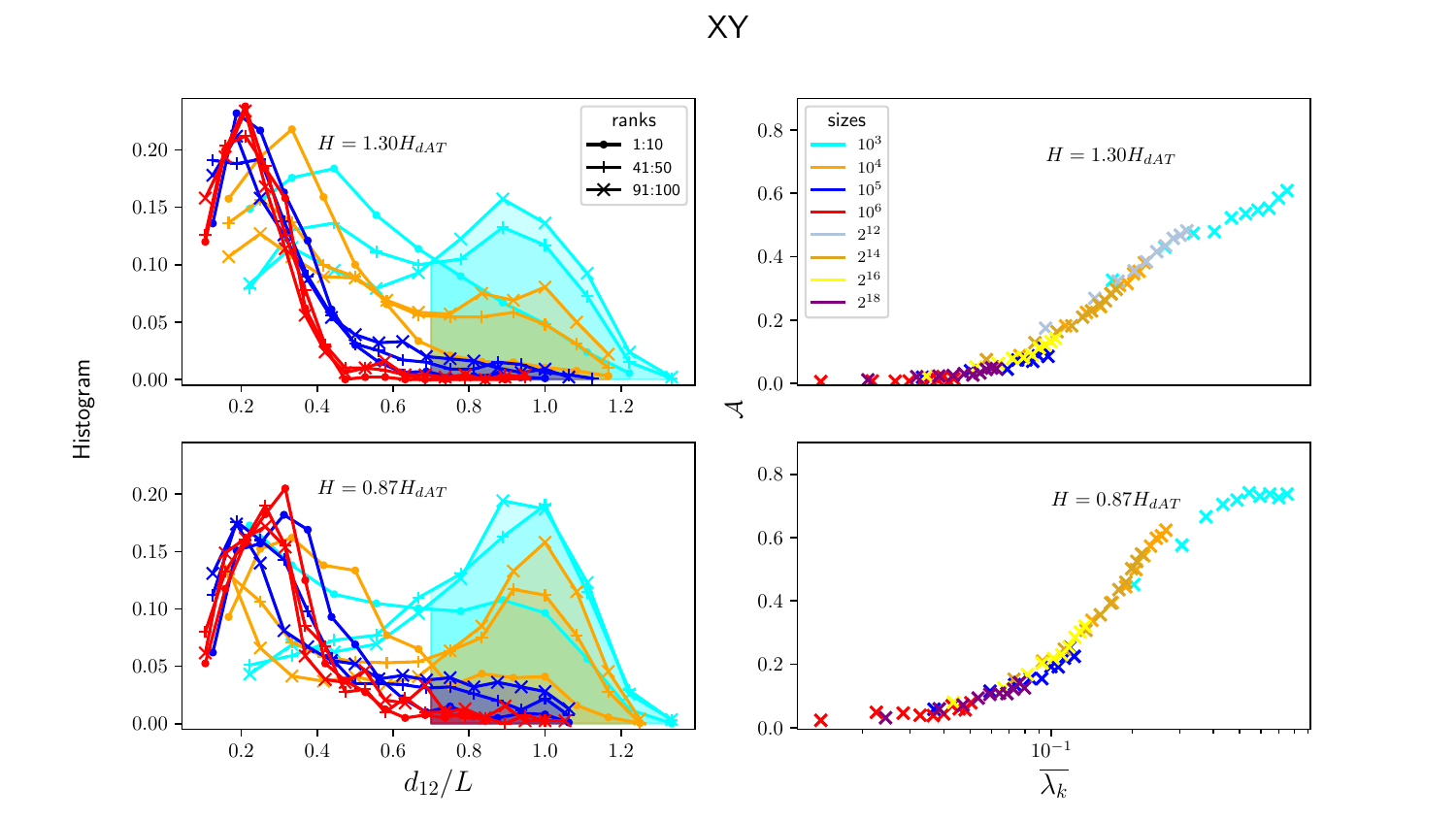}
    \caption{\textbf{Left}: histograms of the distance between the first and second maximum in the WV. Sizes and ranks are indicated in the legends: we aggregated data in intervals $k\in[10 j+1, 10(j+1)]$, with $j=0, \dots, 9$. The colors identifying the different system sizes are the same in all plots.
    \textbf{Right}: the area $\mathcal{A}$ under the histogram for $d_{12}\geq 0.7 L$, as a function of the energy of the excitation $\lambda$. In both cases, $\mathcal{A}^{(k)}$ appears to be a function of $\lambda$. We added data obtained with GOA for sizes $N=2^{12}, 2^{14}, 2^{16}, 2^{18}$.}
    \label{fig:Pgreat_XY}
\end{figure*}

\begin{figure*}
    \centering
    \includegraphics[width=0.75\textwidth]{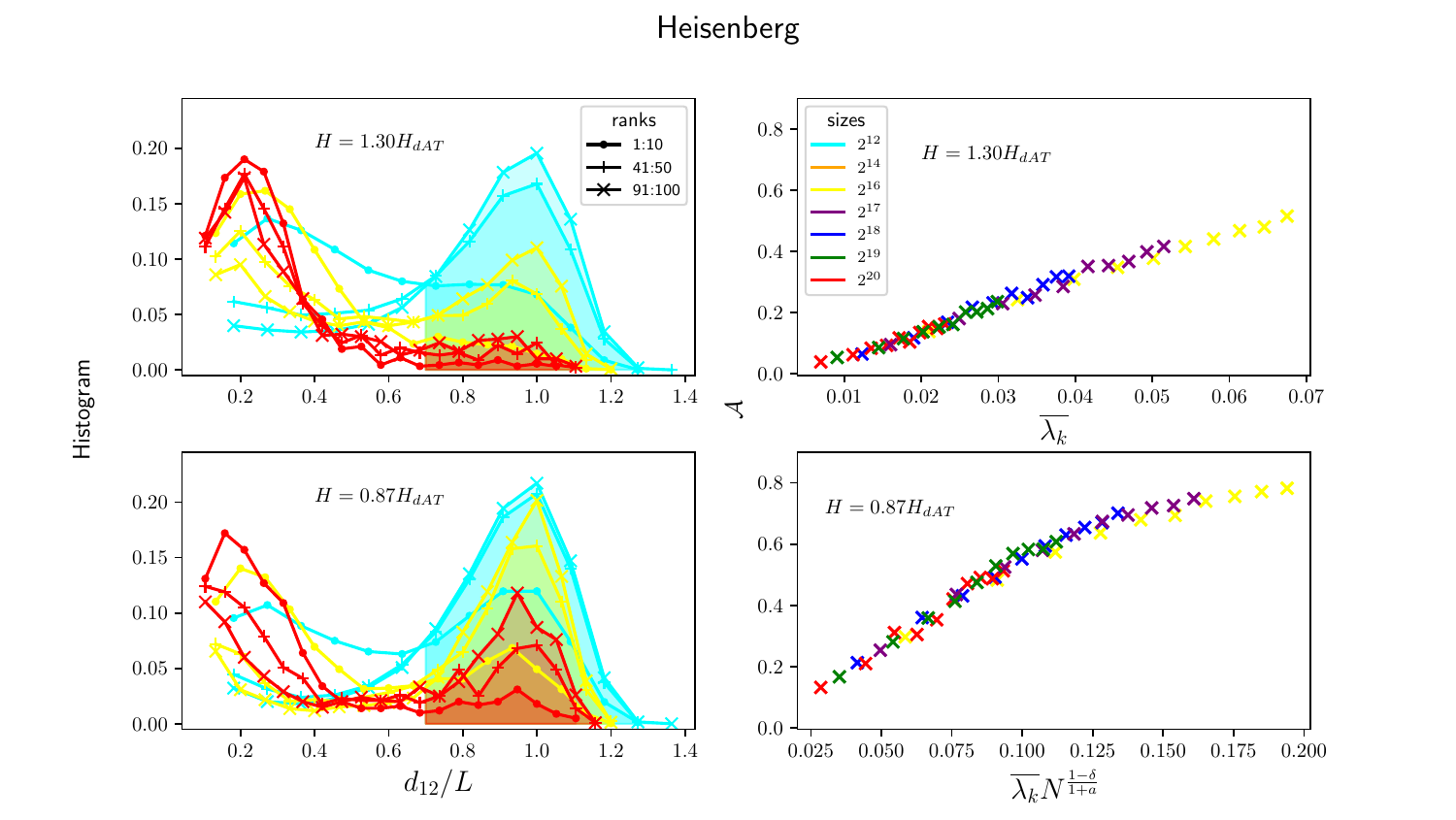}
    \caption{\textbf{Left}: histograms of the distance between principal and secondary maximum in the WV. Sizes and ranks are indicated in the legends. Wehave  aggregated data in intervals $k\in[10 j+1, 10 (j+1)]$, with $j=0, \dots, 9$. The colors identifying the different system sizes are the same in all plots.
    \textbf{Right}: the area $\mathcal{A}$ under the histogram for $d_{12}\geq 0.7 L$, as a function of $\lambda$ for $H=1.30 \Hdat$ and $\lambda N^{-\frac{1-\delta}{1+a}}$ for $H=0.87 \Hdat$, with $\delta\approx 0.65$ roughly for $H=0.87 \Hdat$.}
    \label{fig:Pgreat_Heis}
\end{figure*}

\subsubsection{Measuring probability of CDLEM}

In order to quantify the probability of finding a CDLEM in the spin glass phase of both XY and Heisenberg models we sampled, for each soft mode, the two highest maxima of the WV on the graph, and studied the distribution of their mutual distance. We chose to classify an eigenmode as CDLEM when the distance between the two maxima is larger than a given threshold, $d_{12} = 0.7 L$.
We choose this cutoff based on empirical considerations, as explained below. Note that this is a rather conservative criterion for classifying modes geometrically since we ignored the role of possible other local maxima in the WV. In Appendix~\ref{sec:less_relevant_eigenvector_maxima} we show that our results are not qualitatively affected by considering the contribution from other less dominant peaks.

In Fig.~\ref{fig:visualized_modes}, we provide a visual example of a mode with a single peak (upper panel) and a mode with two distinct peaks (lower panel), in Heisenberg systems of size $N=2^{14}=16384$. The multi-peak modes we wish to study have an IPR that is smaller than that of single-lump modes but many orders of magnitudes larger than $O(1/N)$, the IPR expected for excitations in the bulk of the spectrum.

In the left panels of Figs.~\ref{fig:Pgreat_XY} and~\ref{fig:Pgreat_Heis} we report the histograms of the scaled distances between the two WSD peaks $d_{12}/L$ for any rank $k$.
We observe two distinct populations of modes, identified by two peaks in the histograms: the first at $d=O(1)$ identifies localized excitations, and the second peak at $d \sim L$ identifies CDLEM.
We estimated the probability of observing CDLEM as the area under the histogram curve for $d_{12}/L>0.7$, which is a good cutoff value for discriminating the two populations of modes.

In what follows, we discuss separately the results for the XY and Heisenberg spin glass. For any mode, we consider the statistics of $d_{12}$ against the average eigenvalue $\overline{\lambda_k}$ of the related eigenvector.

\paragraph*{XY.}
Results for the XY model are reported in Fig.~\ref{fig:Pgreat_XY}. The left panels depict the histogram of the distances $d_{12}$, while the right ones report the area under the right peak, that is the probability of a multi-lump mode, $\mathcal{A}^{(k)}$ versus $\overline{\lambda_k}$. The probability $\mathcal{A}^{(k)}$ decreases for increasing sizes and decreasing ranks: more specifically, for fixed $k$, the probability $\mathcal{A}^{(k)}$ vanishes in the infinite-size limit. Therefore, soft modes with rank $k=O(1)$ are localized in this limit. The situation does not qualitatively change when the rank is scaled with powers $N^{\delta}$, with $0<\delta<1$. Indeed, $\mathcal{A}^{(k)}$ appears to be a function of $\overline{\lambda_k}$, both in the paramagnetic and in the spin-glass phase. Low-energy excitations with non-trivial geometry, corresponding to $\mathcal{A}^{(k)}>0$, exist only for $\overline{\lambda_k}>0$ in the infinite-size limit, i.e.\ in the bulk of the spectrum.

\paragraph*{Heisenberg.}
Results for the Heisenberg model are reported in Fig.~\ref{fig:Pgreat_Heis}. As for Fig.~\ref{fig:Pgreat_XY}, left panels show the histogram of the distances $d_{12}$, and the right ones depict the probability $\mathcal{A}^{(k)}$. In the paramagnetic phase, we show this quantity as a function of $\overline{\lambda_k}$, while in the spin-glass phase as a function of the rescaled variable $\overline{\lambda_k} N^{(1-\delta)/(1+a)}$, as in Eq.~\eqref{eq:critical_scale_lambda}.
We found the same exponents $\delta$ as in the study of typical WSD: $\delta \simeq 0.65$ for $H=0.87 \, \Hdat$, and $\delta$ becoming smaller when lowering $H$. Again, such non-trivial scaling implies $\mathcal{A}^{(k)}>0$ for $\lambda\rightarrow 0$, a sign of critical behavior. Low-energy excitations with non-trivial geometry appear at a vanishing energy scale and therefore accumulate to the lower edge in the infinite-size limit.

\section{Conclusions}
\label{sec:concl}

In this work, we have studied numerically the spectrum of the energy Hessian for random-field XY and Heisenberg spin glasses on a random regular graph. We computed low-rank eigenvalues and eigenvectors for excitations around low-lying local energy minima and studied their properties depending on the external field amplitude. These zero-temperature models have a paramagnetic phase for $H>\Hdat$ and a spin-glass phase for $H<\Hdat$: we have studied linear low-energy excitations in both phases.

We have found that both models have a power-law spectral density at small eigenvalues, with an exponent $a=3/2$, which corresponds to an $\omega^4$ behavior in frequency.
Given that we have studied very general models with continuous and bounded variables interacting via a finite connectivity network and non-vanishing interactions, we believe our findings are a strong indication that the $\omega^4$ spectrum is generic for models with disorder (either quenched or self-induced).
Thus our results support the widespread observation of the $\omega^4$ spectrum in glass models \cite{lerner2016statistics, mizuno2017continuum, lerner2017effect, shimada2018anomalous, angelani2018probing, wang2019low, Wang2019sound, richard2020universality, bonfanti2020universal, ji2019, ji2020thermal, ji2021geometry, baity2015soft, Thesis_Lupo2017}, although this has been questioned recently \cite{schirmacher2024nature} for off-lattice glassy systems.
One may question that the models we have studied are defined on a random topology which is rather unrealistic. However, the observation that our results agree with the numerical measurements in the Heisenberg spin glass defined on a three-dimensional cubic lattice \cite{baity2015soft} makes us confident that the random topology is not changing dramatically the physical behavior.

Since the low-energy spectrum does not undergo any particular change at the transition, the spin-glass susceptibility $\chi_{SG}=\langle 1/\lambda^2\rangle$ does not diverge at the transition. The critical behavior of sparse models is therefore qualitatively different from that of dense ones, where instead the zero-temperature transition in a field is signaled by a divergent spin-glass susceptibility \cite{franz2022delocalization}.
The criticality of the model is not visible in the spectrum and must be searched for in the eigenvectors of linear excitations with very low energy.

We have studied the localization properties of soft eigenvectors, both in the paramagnetic and spin-glass phases of the two models.
The first part of the analysis followed the standard approach of identifying through the IPR the number of sites carrying a non-negligible weight of the low-energy eigenmode.
This standard analysis always reports that, in the infinite-size limit, soft modes are localized on a finite number of sites, both in the paramagnetic and in the spin-glass phase. This is at variance with fully-connected vector spin glasses, where soft modes delocalize at the transition, with a vanishing IPR in the large $N$ limit.

Much more informative is the analysis of the spatial extension of the low-energy eigenmodes. Such a geometrical analysis would be completely useless on fully-connected models where every pair of sites are nearest neighbors. Instead on a random graph we can define distances among sites and study the geometrical shape of the subgraph containing most of the weight of a low-energy eigenvector.

The surprising result we have found is that, even with a finite and non-vanishing IPR, that is when the weight is concentrated on a small number of sites, these sites carrying the weight may be spread on very distant regions of the graph.
In this way, the eigenvector gets delocalized even if the susceptible sites are few.
This new kind of delocalization is very different from the standard delocalization where the eigenvectors spread over an extensive number of sites.

So, in addition to the well-known localized phase (weight on few and very close sites) and delocalized phase (weight spread over an extensive number of sites), we are finding a new \emph{concentrated and delocalized} phase where the weight is on few sites spread over very far distances.

While bulk eigenvectors are always delocalized, our interest is in understanding the statistics of low-energy eigenmodes and thus we need to take the large $N$ limit together with the $\lambda \to 0$ limit.
In the paramagnetic phase, as expected, low-energy eigenvectors are always localized (typically around some sites determined by the random field).
In the spin glass phase, the localization properties of the two models differ.
Only in the Heisenberg model, we have found clear indications for a concentrated and delocalized phase for the low-energy eigenmodes. So, in the Heisenberg model, we can connect the spin glass transition with the delocalization of low-energy modes.
Instead, in the XY model, the low-energy modes are still localized: even if they are not as localized as in the paramagnetic phase, our analysis is not able to detect a clear delocalization in the thermodynamic limit on the lower band edge.

Given the key role played by the CDLEM, let us rephrase conclusions in terms of them.
In the paramagnetic phase, there is no CDLEM; in the spin glass phase of the XY model, a CDLEM is rare; in the spin glass phase of the Heisenberg model, a CDLEM is typical.

A CDLEM can be seen as a sort of coexistence between localized and delocalized modes.
Such a coexistence was observed in other models in different situations: for instance, in~\cite{das2023absence} is observed for a 1D Anderson model with heterogeneous hoppings, for bulk eigenvectors around zero energy.

The existence of CDLEM is a clear sign of criticality.
Indeed it corresponds to highly susceptible sites placed at arbitrarily large distances (as in a critical model).
Compared to their typicality in the Heisenberg spin glass phase, the rareness of CDLEM in the XY model is possibly due to the different nature of the degrees of freedom, or to the unavoidable limitations of our numerical investigation. Using the Arnoldi exact diagonalization, we computed only the lowest 100 eigenmodes: if a signal of critical behavior is present, it could be detectable at a scale $100 \ll k\ll N$, hardly accessible through our method.

We believe that a series of follow-up studies are needed to fully understand the delocalization transition in sparse models.
Firstly, the analysis of non-linear excitations. The exponent $a=3/2$ we found is compatible with the prediction of the Soft Potential Model \cite{gurarie2003bosonic}. To understand if our theory is consistent with it, a future line of research would be to study non-linear soft modes around local minima of the energy, following for instance the method described in~\cite{lerner2021low}, to detect possible double-well excitations and study their relations with soft eigenmodes. The behavior of soft non-linear excitations could be more informative on the nature of the transition.

Secondly, the Bethe lattice theory. The results found in this work for the spectrum come from numerical simulations. A necessary follow-up of our work would be to study the spectrum through the cavity approach, by solving the cavity resolvent equations~\cite{AbouChacraEtAl1973, Vivo2021lectures}. This method would allow us to access the spectrum beyond the lower edge, so it is ideal to study the behavior of excitations at different energy scales.
To our knowledge, a systematic study of the spectrum trough the cavity approach has not yet been carried out for sparse vector spin glasses. By solving the Bethe lattice theory for the spectrum, we expect to identify the correct susceptibility function for the zero-temperature spin-glass transition of sparse vector spin glasses.

\begin{acknowledgments}
    The research has received financial support from ICSC -- Italian Research Center on High-Performance Computing, Big Data, and Quantum Computing, funded by the European Union -- NextGenerationEU and from the Simons foundation (grants No. 454941, S. Franz and No. 454949, G. Parisi).
\end{acknowledgments}

\bibliographystyle{apsrev4-1}
\bibliography{myBiblio}

\clearpage

\onecolumngrid
\appendix

\renewcommand{\thefigure}{S\arabic{figure}}
\renewcommand{\theequation}{S\arabic{equation}}
\setcounter{figure}{0}
\setcounter{section}{0}
\setcounter{equation}{0}

\title{Supplementary Information}

\section{The gapped phase}
\label{sec:gapped_phase}

Here we discuss the  gapped spectrum phase of the models studied by us in this work.

\subsection{The gap of the local fields distribution}

\begin{figure*}[h]
    \centering
    \includegraphics[width=0.7\textwidth]{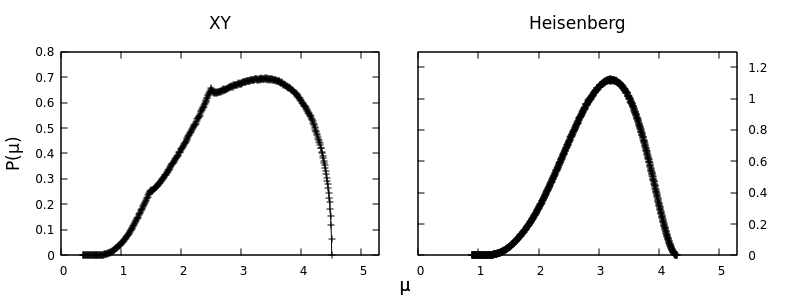}
    \caption{PDFs of the local fields strengths $\mu_i$ for $H=1.30 \Hdat<c=3$. The gap at low fields is clearly visible.}
    \label{fig:loc_fields_gap}
\end{figure*}

The gap in the spectrum stems from the local fields magnitudes $\mu_i\equiv |\boldsymbol{\mu}_i|$ (the local fields are defined by eq. \eqref{eq:local_field}): when the external field amplitude is very strong, $\mu_i\simeq H+O(1)$ and the interactions cannot induce soft excitations in the system. 
By purely geometrical considerations, one can find the following bounds on the $\mu_i$ of our model, for $H>c$: 
\begin{eqnarray}
\label{eq:loc_field_bounds}
    & H-c\leq \mu_i \leq H+c\qquad H>c 
\end{eqnarray}
However, the gap in the local field distribution is not present just for $H>c$. In fact, due to the well known phenomenon of Onsager polarization \cite{Palmer_1979}, the magnitudes of local fields has to be strictly positive in order for the system to be stable: for any site $i$, 
$ \mu_i>0 $.
Let us briefly show how to deduce, within the cavity approach, the necessary presence of a gap in the distribution of the local fields strengths $\mu_i$. The local magnetic susceptibilities $\chi_{ii}^{\alpha\beta}$ satisfy the spectral cavity equations
\begin{equation}
\label{eq:local_susc}
    \chi_{ii}^{\alpha\beta}=(\mu_i\boldsymbol{I}-\sum_{j\in\partial i}J_{ij}^2\boldsymbol{P}_i^{\perp}\boldsymbol{{\chi}}_{j\rightarrow i}\boldsymbol{P}_j^{\perp})^{-1}_{\alpha\beta}
\end{equation}
where $\boldsymbol{\chi}_{j\rightarrow i}$ are the cavity susceptibilities, i.e. the susceptibilities on a graph where site $i$ and its links have been removed, and $\boldsymbol{P}_i^{\perp}$ are orthogonal projectors with respect to spin $\boldsymbol{S}_i$. The local susceptibilities are well defined if and only if the local susceptibility matrix is positive defined, i.e. 
\begin{equation}
    \operatorname{det}(\mu_i\boldsymbol{I}-\sum_{j\in\partial i}J_{ij}^2\boldsymbol{P}_i^{\perp}\boldsymbol{{\chi}}_{j\rightarrow i}\boldsymbol{P}_j^{\perp})^{-1} > 0
\end{equation}
This condition imposes that local fields $\mu_i$ cannot be arbitrary small. 
In the dense case $c=O(N)$ the Onsager reaction term is the same on all sites \cite{Palmer_1979}: for any $i$ one has simply $\mu_i>\chi$. In fig \ref{fig:loc_fields_gap} we show the distribution of local fields for $H=1.30 H_{dAT}<c=3$ for both the XY and Heisenberg model. In both cases, it is evident that the distribution of local fields is gapped.

\subsection{The spectral gap and gap closure}

The gapless spectrum phase emerges when interactions become relevant for the statistics of the $\mu_i$.
Let us show how to deduce an upper bound for the external field at which the spectrum becomes gapless, be it $H=H_{gap}$.
We make us of the eigenvalue equation
\begin{equation}
    \label{eq:specEQ}
    [(\boldsymbol{\mathcal{M}}-\lambda\mathbf{I})\boldsymbol{\underline{\psi}}]_i=-\sum_{j\in\partial i}J_{ij}\mathbb{P}^{(\perp)}({\bf S}_i)\boldsymbol{\psi}_j(\lambda)+(\mu_i-\lambda)\boldsymbol{\psi}_i=\boldsymbol{0}
\end{equation}
where $\boldsymbol{\underline{\psi}}(\lambda)\equiv (\boldsymbol{\psi}_1,\dots,\boldsymbol{\psi}_N)$ is an eigenvector and each component $\boldsymbol{\psi}_i$ a $m$-dimensional vector and $\mathbb{P}^{(\perp)}(\bf{S}_i)=\mathbb{I}-\bf{S}_i\bf{S}_i^T$ is a projector on the space orthogonal to $\bf{S}_i$. By doing the scalar product of both hands of \eqref{eq:specEQ} with $\boldsymbol{\psi}_i$, we can rewrite \eqref{eq:specEQ} as
\begin{equation}
    \lambda\,=\,\mu_i-\sum_{j\in\partial i}J_{ij}\frac{\mathbb{P}^{(\perp)}(\bf{S}_i)\boldsymbol{\psi}_j\cdot \boldsymbol{\psi}_i}{\psi_i^2}\,=\,\mu_i-\sum_{j\in\partial i} J_{ij}\frac{\psi_j}{\psi_i}(\hat{\boldsymbol{\psi}}_i\cdot\hat{\boldsymbol{\psi}}_j)
\end{equation}
where $\hat{\boldsymbol{\psi}}_i$ are unit direction vectors.
We can then write a lower bound for $\lambda$
\begin{eqnarray}
\label{eq:lower_bound_eigs}
    && \lambda=\mu_i-\sum_{j\in\partial i}J_{ij}\frac{\mathbb{P}^{(\perp)}(\bf{S}_i)\boldsymbol{\psi}_j\cdot \boldsymbol{\psi}_i}{\psi_i^2}>H-c-K\,\equiv\,\lambda_{gap}^{(0)} \\
    && K\,\equiv\,\max_i \sum_{j\in\partial i} J_{ij}\frac{\psi_j}{\psi_i}(\hat{\boldsymbol{\psi}}_i\cdot\hat{\boldsymbol{\psi}}_j)
\end{eqnarray}
where with $\lambda_{gap}^{(0)}$ we mean an estimate of the spectral gap using the rough lower bound $\mu_{gap}\,=\,H-c$ for the local fields, i.e. ignoring $O(1/H)$ corrections that might come from Onsager reaction. 
The gap external field is found by imposing $\lambda_{gap}^{(0)}=0$ in \eqref{eq:lower_bound_eigs}:
\begin{equation}
\label{eq:field_gap}
    H_{gap}^{(0)}\,=\,c+K.
\end{equation}
The quantity $K$ can be evaluated through the following considerations. Given that $i$ is the site where the eigenvector is strongly localised, one can expect for the eigenvector components on its neighbors $\psi_j\sim \epsilon_j\psi_i$, for some $0<\epsilon_j<1$: the simplest choice is to set $\epsilon_j\sim 1/\sqrt{c}$ giving $K=\sqrt{c}$ and eq. \eqref{eq:field_gap} returns for $c=3$
\begin{equation}
\label{eq:field_gap_order_zero}
    H_{gap}^{(0)}\,\simeq\,c+\sqrt{c}=4.73205081\dots.
\end{equation}

\subsection{Numerical measures of the spectral gap}

\begin{figure*}
    \centering
    \includegraphics[width=0.4\textwidth]{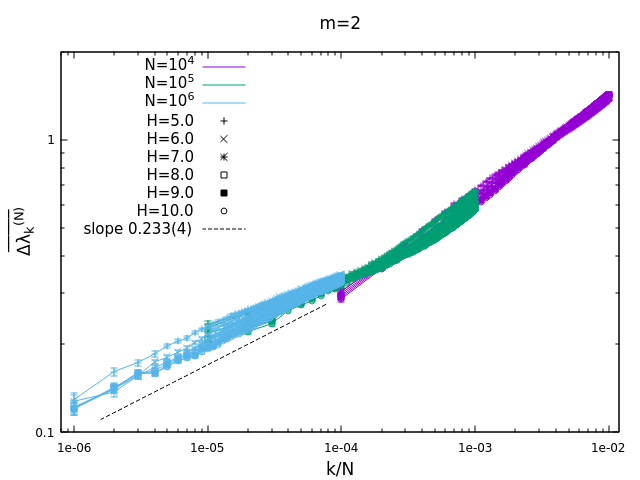}
    \includegraphics[width=0.4\textwidth]{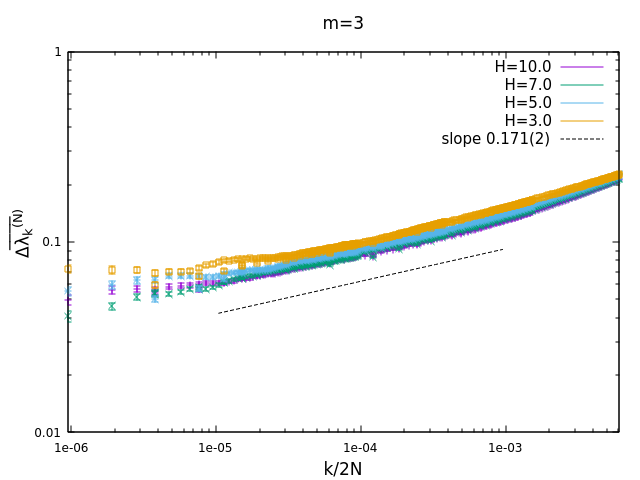}
    \caption{The average $\Delta\lambda_k^{(N)}$ versus the rescaled ranks, on the left the XY model and on the right the Heisenberg one. The estimate of the power law exponent at low rescaled rank allows for a measure of the lower spectral edge.}
    \label{fig:deltaLambda}
\end{figure*}

\begin{figure*}
    \centering
    \includegraphics[width=0.4\textwidth]{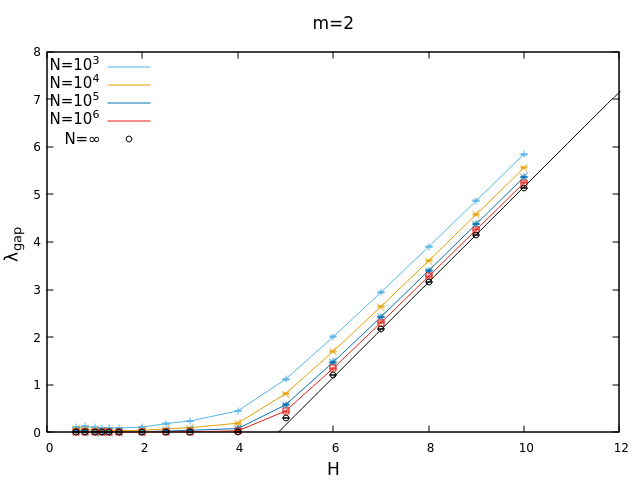}
    \includegraphics[width=0.4\textwidth]{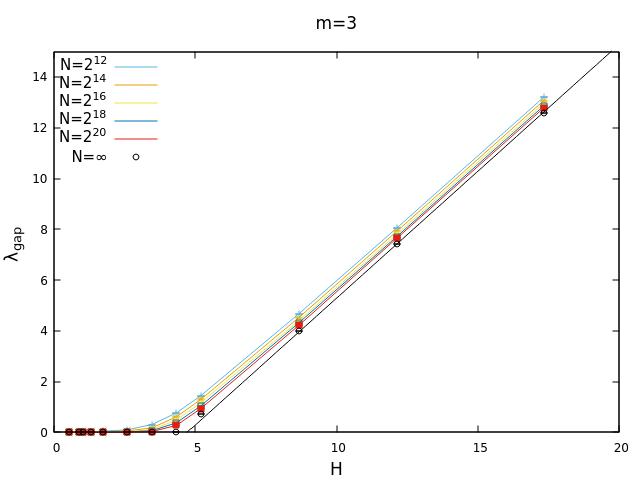}
    \caption{The lower spectral edge, for finite sizes and in the thermodynamic limit, as estimated through \eqref{eq:lower_edge_gap}, as a function of $H$. On the left the XY model, on the right the Heisenerg one. The two graphs are qualitatively identical, with an upper bound for the gap-closure field sligthly different between the two models. We show, as a visual reference, the continuations of the curves in the gapless phase.}
    \label{fig:lambdaGAP}
\end{figure*}

We measured numerically the spectral gap $\lambda_{-}$ for $H>H_{gap}^{(0)}$, in order to check whether our estimate \eqref{eq:field_gap_order_zero} is correct or not: calling $k$ the rank of the modes, we exploited the equation for sample-averaged lower edge eigenvalues
\begin{equation}
\label{eq:lower_edge_gap}
    \overline{\lambda_k^{(N)}}\sim \lambda_{-}+A\left(\frac{k}{N}\right)^{\frac{1}{1+a}}
\end{equation}
and considered the differences
\begin{equation}
    \overline{\Delta\lambda_k^{(N)}}\equiv \overline{\lambda_k^{(N)}}-\overline{\lambda_k^{(l N)}}\sim A\left(1-\frac{1}{l^{1/(1+a)}}\right)\left(\frac{k}{N}\right)^{\frac{1}{1+a}}
\end{equation}
using a fixed $l$ for each couple of sizes (we used $l=10$ in the XY case and $l=4$ in the Heisenberg case).
The exponent $a$ can be measured simply through a linear fit of $\log \overline{\Delta\lambda_k^{(N)}}$ versus $\log(k/N)$. The spectral gap $\lambda_{-}$ is then obtained by extrapolation from \eqref{eq:lower_edge_gap}. In figure \ref{fig:deltaLambda} we show our measures of $\overline{\Delta\lambda_k^{(N)}}$ versus $k/N$, with our estimates of the exponent $a$: we found
\begin{eqnarray*}
    & a=3.30(10)\qquad XY \\
    & a=4.85(13)\qquad Heisenberg.
\end{eqnarray*}
In figures \ref{fig:lambdaGAP} we show our numerical measures of the spectral gap as a function of $H$. By extrapolating using data points with the largest $H$, in the two cases we found the gap field upper bounds
\begin{eqnarray*}
    & H_{gap}^{(0)}=4.85(3)\qquad XY \\
    & H_{gap}^{(0)}=4.71(2)\qquad Heisenberg
\end{eqnarray*}
Only the estimate in the Heisenberg case seems to be compatible with the guess $H_{gap}^{(0)}=3+\sqrt{3}$. In any case, our data points of the spectral gap deviate from the large $H$ linear extrapolation close to it, suggesting that the true field at which the spectrum becomes gapless is at a value lower than our estimated upper bounds.

\section{Robustness of the spectral exponent}

\label{sec:robustness_spectral_exponent}

Here we analyse the robustness of our measures of the exponent $a$ of the lower tail of the spectral density, against the degree of GOA used; we restrict our analysis to the $m=3$ case. In addition, we test the robustness of exponent $a$ against the dimensions of spins $m$, studying the case of a spin glass with $m=4$.

\subsection{Minimisation through OR}
\begin{figure}[h]
    \centering
    \includegraphics[width=0.35\textwidth]{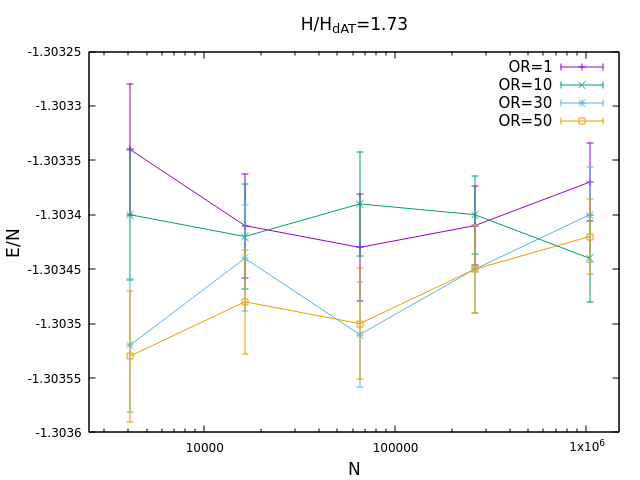}
    \includegraphics[width=0.35\textwidth]{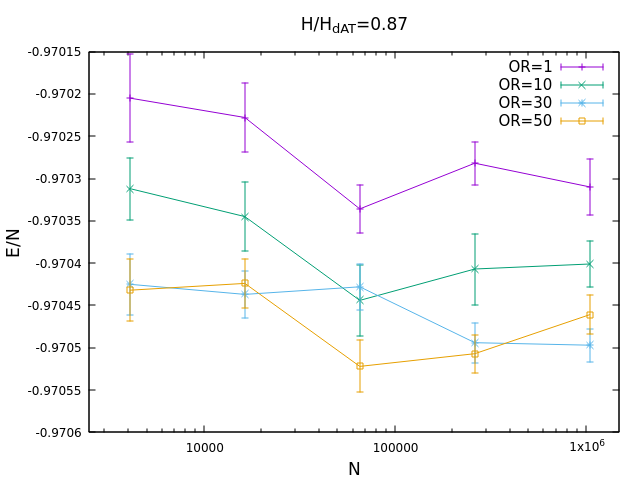}
    \caption{The average energy density of minima reached by minimisation with the GD-OR \eqref{eq:GD_with_OR}, for different over-relaxations. Higher values of $\mathcal{O}$ are associated to deeper minima in the landscape. In the paramagnetic phase there is no significant difference in performance with different choices of the OR, while in the spin glass phase a sufficiently high value of GOA is necessary to avoid low quality energy minima.}
    \label{fig:energy_OR}
\end{figure}

We begin to study how the measure of $a$ is affected by the Over-Relaxation (OR) used in our minimisation algorithm. Since we mostly used this algorithm in the $m=3$ case, we restrict our analysis to this case. We consider only two values of field amplitude, $H=1.73 \Hdat$ and $H=0.87 \Hdat$, as representatives of the paramagnetic and spin glass phases respectively. The Over-Relaxation (OR) minimisation algorithm is described by eq. \eqref{eq:GD_with_OR}. We found that deeper minima in the landscape are reached the higher is the value of the OR: we show this in figure \ref{fig:energy_OR}, where we plot the average energy density $E/N$ reached in the minimisation as a function of the size of the system, for different over-relaxations. The difference in the average energy reached in the minimisation is significant only for $H<H_{dAT}$.

\begin{figure}[h]
    \centering
    \includegraphics[width=0.35\textwidth]{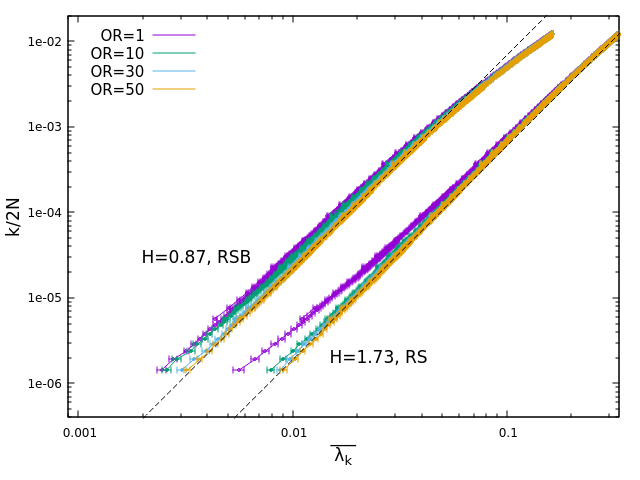}
    \includegraphics[width=0.35\textwidth]{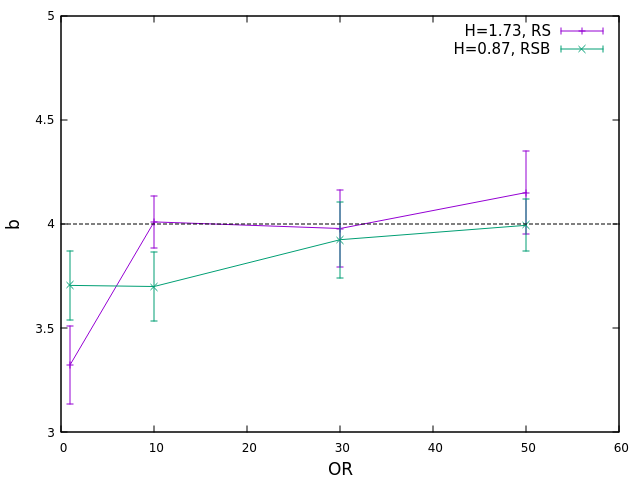}
    \caption{Left: the empirical CDF of eigenvalues, for $H/\Hdat=0.87, 1.73$, for different values of the over-relaxation parameter. The spectra of minima with lower $\mathcal{O}$, which are related to higher energy minima as shown in the bottom figures, are richer of low energy excitations.\newline Right: the exponent of the DoS $b=2a+1$, where $a$ is the exponent of the spectal density. If $\mathcal{O}$ is not high enough, high minima of the energetic landscape are reached and $3<b<4$. For $H/\Hdat=1.73$, $b=4$ for $\mathcal{O}\geq 10$, while $\mathcal{O}\geq 30$ for $H/\Hdat=0.87$.}
    \label{fig:robustness_minimis}
\end{figure}

We found that the lower tail exponent of the DOS $b=2a+1$ is $b=4$ when the over-relaxation parameter $\mathcal{O}$ of the GD-OR algorithm is sufficiently high, i.e. where minima deep enough in the landscape are reached. In the left plot of figure \ref{fig:robustness_minimis} we show for the external fields $H=0.87, 1.73$, referring respectively to the RSB and the RS phase, the lower tails of the empirical cumulative distributions of eigenvalues, showing a curve for each value of $\mathcal{O}=1, 10, 30, 50$. The slope of the tail in this log-log plot seems to be lower for small OR, suggesting as a consequence a value $b<4$ for the exponent of the DoS.
This observation is substantiated by the right plot of figure \ref{fig:robustness_minimis}, which shows measures of the exponent $b$ of the DOS as a function of the OR: we see that for sufficient large $\mathcal{O}$ the measures are consistent with $b=4$ within one error bar. The value of $\mathcal{O}$ necessary to robustly measure such a value increases as $H$ is decreased: in the RS case given by $H=1.73 \Hdat$ we find $b=4$ for $OR=10, 50$, while in the RSB case $H=0.87 \Hdat$ we observe $b=4$ only for $\mathcal{O}=50$. This result suggests that it is increasingly harder to reach good low energy minima as $H$ is decreased, therefore a higher degree of GOA is required the deeper one goes in the spin glass phase.

The fact that the statistics of the spectral lower edge is different for high energy minima is consistent with the double-well picture portrayed by the soft potential model. In a random quartic polynomial, the high well is on average flatter than the lower one: the distribution of small oscillations frequencies is cubic in the high well and quartic in the low one \cite{GurarieChalker2003, rainone2021mean}.

\subsection{The $m=4$ spin glass}

\begin{figure}[h]
    \centering
    \includegraphics[width=0.4\textwidth]{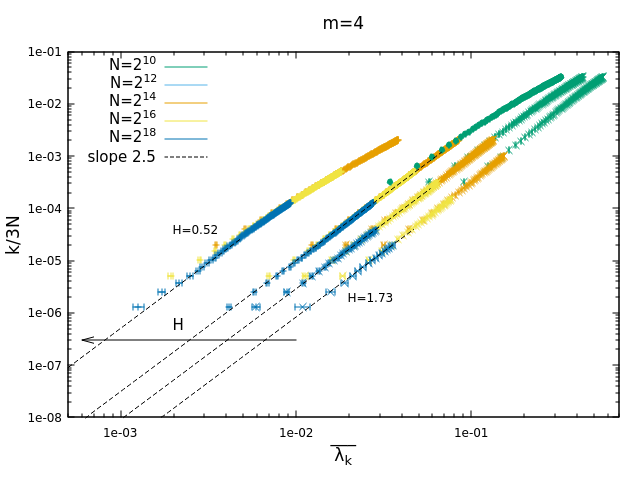}
    \includegraphics[width=0.4\textwidth]{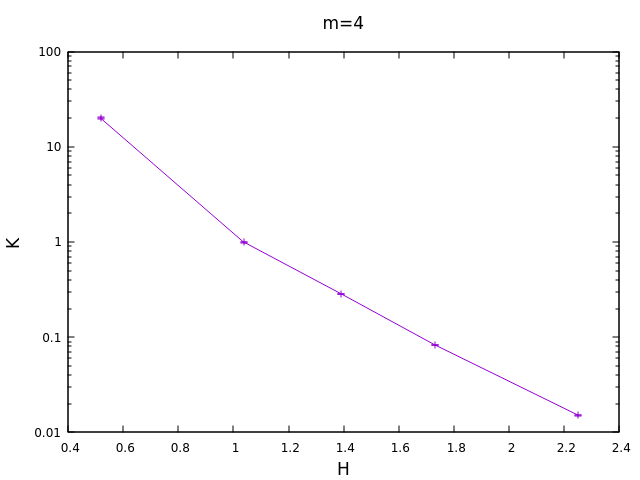}
    \caption{Left: the lower tail of the empirical cumulative distribution of eigenvalues of the $m=4$ spin glass. For several values of $H$, the lower tail exponent is consistent with the value $2.5$ and thus with a quartic law for the DoS. Right: the spectral prefactor as a function of $H$. The qualitative behavior of the system is identical to that of the $m=3$ model.}
    \label{fig:m=4_avgCumul}
\end{figure}

We show a few results of the system with four-dimensional spin variables. We considered a RRG with $c=3$ and the same distribution of disorder for the couplings and external fields as those used in the work.
The properties of this spin glass are qualitatively identical to that of the Heisenberg ($m=3$) one. We begin by showing our measures of the lower tail exponent $a$ and the spectral prefactor $K$ featuring in the CDF, which we recall the reader reads
\begin{equation*}
    F(\lambda)\sim K\lambda^{1+a}
\end{equation*}
close to the lower edge.
In figure \ref{fig:m=4_avgCumul} we show on the left plot curves of the cumulative distributions of eigenvalues, for values $H=1.73, 1.39, 1.04, 0.52$. Our numerical data are consistent with a quartic law for the DOS, for a wide range of values of $H$. Analogously to the Heisenberg $m=3$ case, the prefactor is a monotonic function of $H$ and spans several order of magnitudes in the range of fields shown in the right plot of \ref{fig:m=4_avgCumul}. Thus, the system gains more and more low energy modes as the field is lowered. No estimate of the critical field of this model is known, to our knowledge. It appears to us that the physics of the $m=4$ spin glass is qualitatively similar to that of the $m=3$ one.

\section{Robustness of eigenvector statistical properties against over-relaxation}
\label{sec:robustness_eigenvectors}

\begin{figure}[h]
    \centering
    \includegraphics[width=0.4\textwidth]{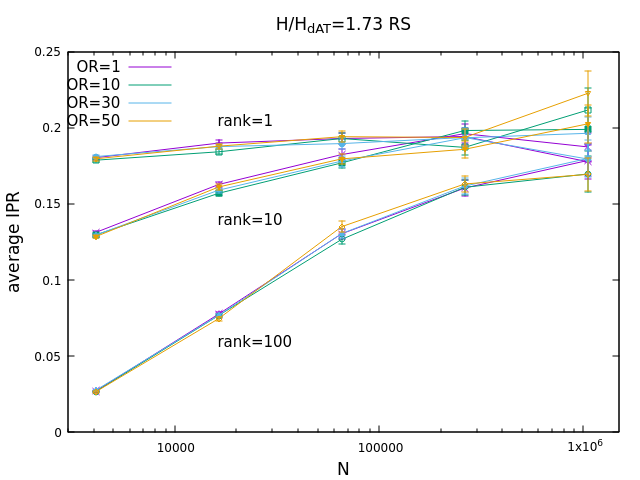}
     \includegraphics[width=0.4\textwidth]{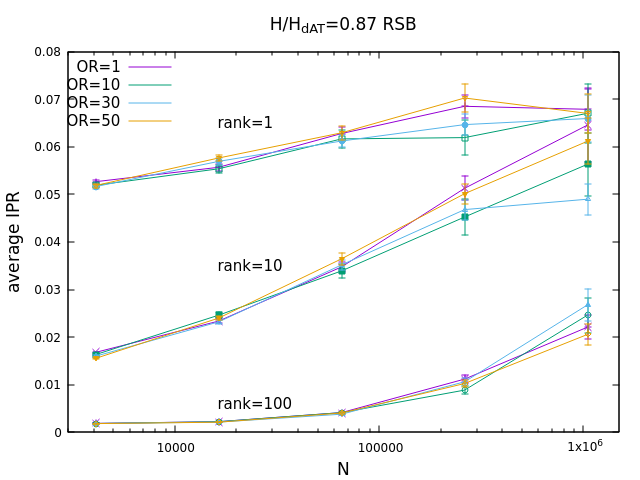}
    \caption{Average IPR versus size $N$ curves, obtained from simulations with different OR, as explained in the legend. On the left simulations in the RS phase, on the right in the RSB phase. In both cases, the average IPRs do not depend on the OR: all data points are compatible within two errorbars.}
    \label{fig:robustness_IPR}
\end{figure}

We found that the statistical properties we observe for the eigenvectors are essentially independent from the energy minima reached through over-relaxation. We show results in the Heisenberg case.
In figure \ref{fig:robustness_IPR} we show the IPR versus $N$, for values of over-relaxation $\mathcal{O}=1, 10, 30, 50$. Both in the paramagnetic phase (left plot) and in the spin glass phase (right plot), the sample average IPR does not depend on the value of OR, and thus on the range of energy levels reached by minimisation through OR. Larger fluctuations are observed for the larger sizes, even though their magnitude is compatible with single measures error bars.

\begin{figure}[h]
    \centering
    \includegraphics[width=0.4\textwidth]{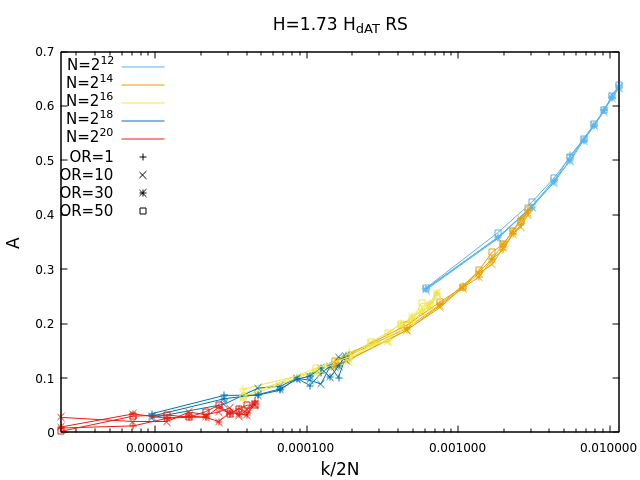}
     \includegraphics[width=0.4\textwidth]{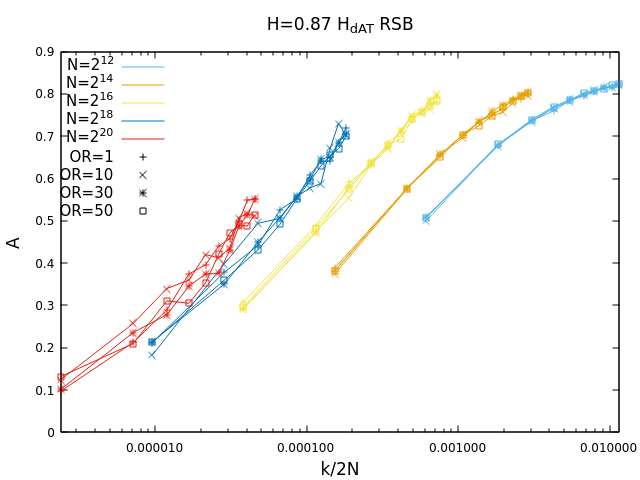}
    \caption{The proability of observing multi-peaked eigenmodes versus the rescaled rank $k/2N$, for different values of OR, as explained in the legend. The dependence on the OR is very weak, being manifest only for the largest size $N=2^{20}$.}
    \label{fig:robustness_Pgreat}
\end{figure}

In order to provide more evidence of the robustness of eigenvectors properties, we also consider their geometry on the graph.
In the plots of figure \ref{fig:robustness_Pgreat} we show the probability of observing CDLEM (Concentrated and Delocalised Low Energy Modes), which in the plots is the quantity $\mathcal{A}$, versus the relative rank $k/2N$, for different values of OR. For any size, curves related to different levels of OR overlap very well, suggesting that our numerical observations of the frequency of CDLEM are robust.

\section{Statistics of eigenvalues spacings}

\label{sec:r}

An alternative way to understand the spectral properties of the system is to study the statistics of eigenvalues spacings across different energy scales. We measured for all sizes and external fields the ratio
\begin{equation}
\label{eq:r_def}
    r_i\,=\,\frac{\min(\Delta\lambda_i, \Delta\lambda_{i+1})}{\max(\Delta\lambda_i, \Delta\lambda_{i+1})}
\end{equation}
where $\Delta\lambda_i=\lambda_{i+1}-\lambda_i$ and $i=1, \dots, 98$. These parameters where firstly introduced in \cite{oganesyan2007localization} to study the local statistical properties of the spectrum of large quantum Hamiltonians. The statistics of the parameters in \eqref{eq:r_def} bear a more robust description of the local spectral properties: in fact, the distribution of level spacings depends on the local density of states, whereas that of the $r$ does not. In particular, $r$ is more effective to describe the statistics of level spacings approaching a spectral edge with vanishing density of states, as it is our case of study.  
The connection between the statistics of level spacings and that of the $r_i$ is the following: in the case of Wigner-Dyson statistics, the distribution of $r$ is approximately
\begin{equation}
\label{eq:pdf_r_wigner_surmise}
    P_w(r)\,=\,\frac{27}{4}\frac{r+r^2}{(1+r+r^2)^{5/2}}\theta(1-r)\theta(r)
\end{equation}
as derived in \cite{atas2013distribution}, starting from the Wigner surmise 
\begin{equation*}
    P(\Delta\lambda)=A\;\Delta\lambda\;e^{-\frac{\Delta\lambda^2}{2}}
\end{equation*}
which approximates the true distribution of spacings in the Gaussian Orthogonal Ensemble (GOE). This spacings statistics usually identifies delocalised modes in the bulk of the spectrum of classical random matrix ensembles.  
At variance, localised modes are usually identified by their eigenvalues following a Poisson point process, that is, with the $\Delta\lambda_i$ being distributed exponentially. In this case, it is easy to show that
\begin{equation}
\label{eq:pdf_r_poisson}
    P_p(r)=\frac{2}{(1+r)^2}\theta(1-r)\theta(r)
\end{equation}
The pdf of the parameter $r$ not necessarily is of one of the two forms \eqref{eq:pdf_r_wigner_surmise}, \eqref{eq:pdf_r_poisson}: for fractal or multifractal delocalised modes, usually the pdf of $r$ has an intermediate functional form. In particular, the expected value $\overline{r}$ satisfies
\begin{eqnarray}
    & \mathrm{r}_{p}\leq \overline{r}\leq \mathrm{r}_{w}
\end{eqnarray}
where 
\begin{equation}
    \mathrm{r}_{p}=\int\;dr\;P_p(r)r=2\log2-1\approx 0.38629
\end{equation}
\begin{equation}
    \mathrm{r}_{w}=\int\;dr\;P_w(r)r=4-2\sqrt{3}\approx 0.53590.
\end{equation}
The value of $\mathrm{r}_{w}$ obtained just above is an approximation: the first moment in the GOE case was measured numerically by
Atas et al. \cite{atas2013distribution}, obtaining $\tilde{\mathrm{r}}_{w}=0.5307(1)$: we will use this upper bound in our numerical measures. 

\begin{figure*}[h]
    \centering
    \includegraphics[width=0.45\textwidth]{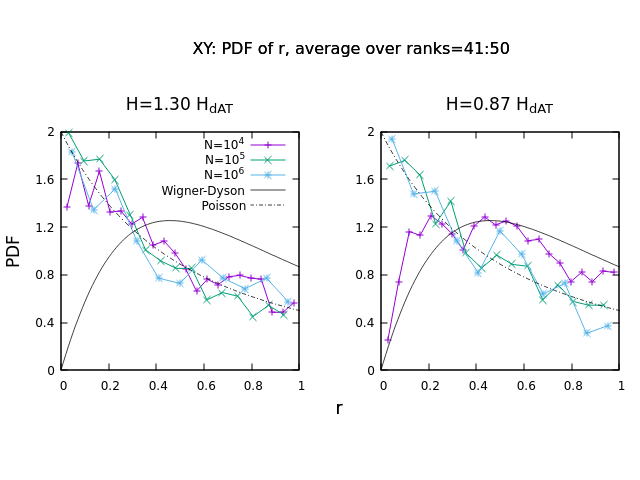}
    \includegraphics[width=0.45\textwidth]{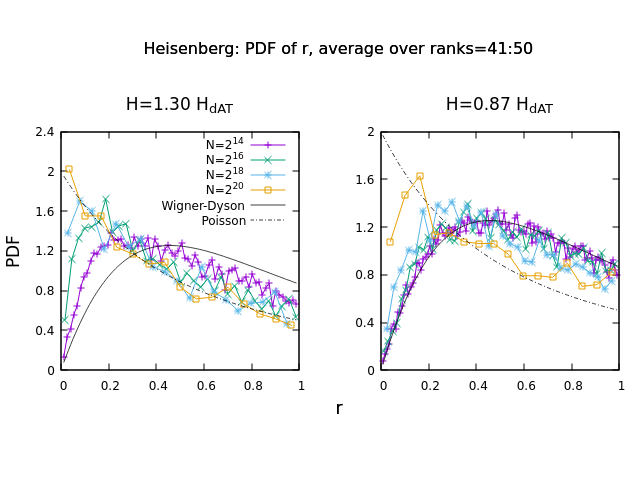}
    \caption{\textbf{Left}: the pdf of $r$ in the XY model, comparing values of $H>H_{dAT}$ and $H<H_{dAT}$. Here, we keep the ranks fixed for increasing sizes, showing PDF obtained by data aggregated in a window of modes with ranks $k=41:50$.
    \newline
    \textbf{Right}: the pdf of $r$ in the Heisenberg model, comparing values of $H>H_{dAT}$ and $H<H_{dAT}$. Here, we keep the ranks fixed for increasing sizes, showing PDF obtained by data aggregated in a window of modes with ranks $k=41:50$.
    }
    \label{fig:pdf_r_fixed_rank}
\end{figure*}

\begin{figure*}[h]
    \centering
    \includegraphics[width=0.45\textwidth]{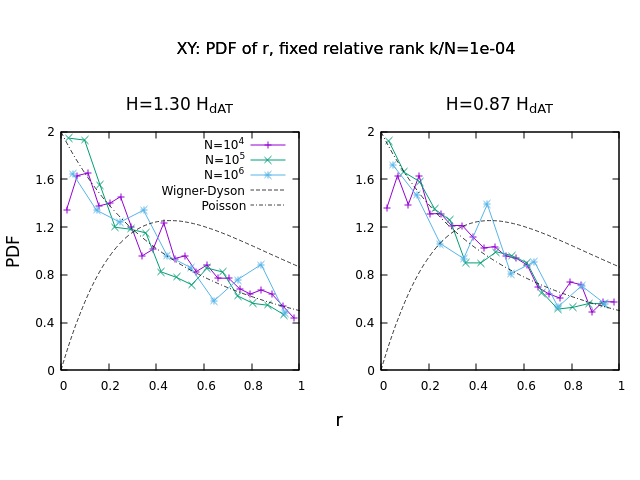}
    \includegraphics[width=0.45\textwidth]{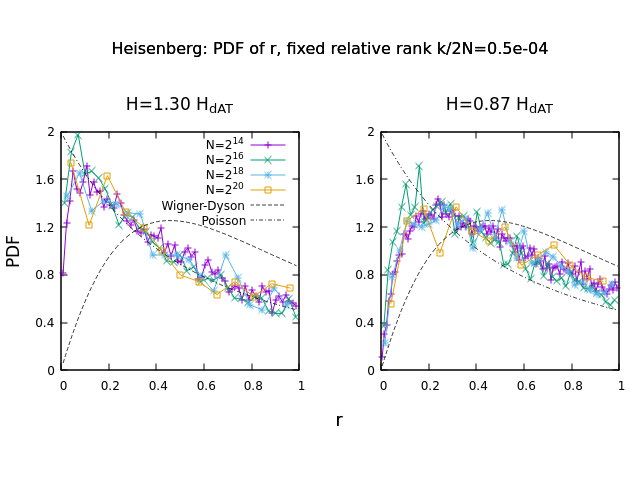}
    \caption{\textbf{Left}: the pdf of $r$ in the XY model, comparing values of $H>H_{dAT}$ and $H<H_{dAT}$. Here, we keep the relative rank fixed to $\kappa=10^{-4}/(m-1)$, showing PDF obtained by data aggregated in windows of ten modes.
    \newline
    \textbf{Right}: the pdf of $r$ in the Heisenberg model, comparing values of $H>H_{dAT}$ and $H<H_{dAT}$. Here, we keep the relative rank fixed to $\kappa=10^{-4}/(m-1)$, showing PDF obtained by data aggregated in windows of ten modes.}
    \label{fig:pdf_r_fixed_relative_rank}
\end{figure*}

\begin{figure*}[h]
    \centering
    \includegraphics[width=0.45\textwidth]{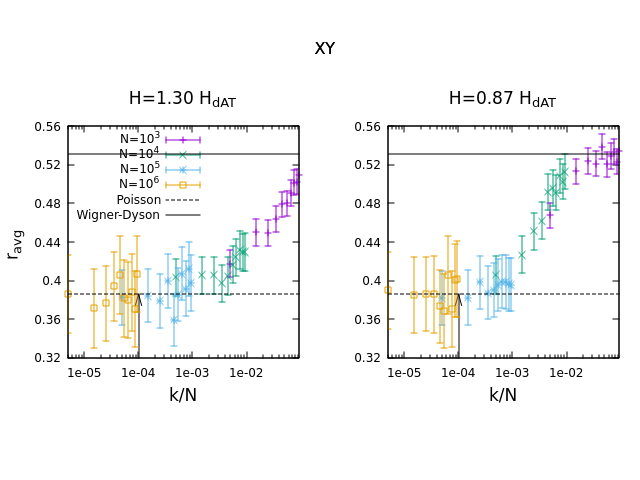}
    \includegraphics[width=0.45\textwidth]{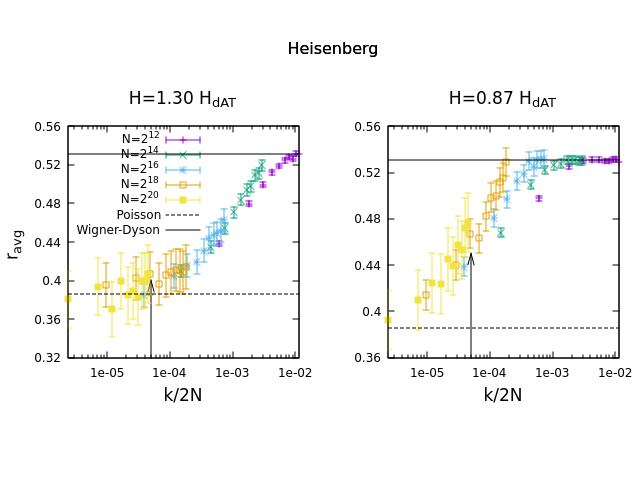}
    \caption{The average value of $r$ as a function of relative rank. With a full black arrow, we show the value of $\overline{r}$ related to the value $\kappa=10^{-4}/(m-1)$ used in the previous figure.
    \newline
    \textbf{XY}: at low relative rank the curves related respectively to the paramagnetic and the spin glass phases are not significantly different. 
    \newline
    \textbf{Heisenberg}: at low relative rank the curves related respectively to the paramagnetic and the spin glass phases are significantly different. 
    }
    \label{fig:r_avg}
\end{figure*}

In the following, we convey our results. In figure \ref{fig:pdf_r_fixed_rank} we show the pdf of $r$, comparing the paramagnetic and spin glass phase in the XY (left) and Heisenberg (right) models respectively. We collect eigenvalues in a fixed rank window $41\leq k \leq 50$ for varying sizes, in order to enlarge the size of the statistical samples. The pdf of $r$ is close to the Wigner surmise for small sizes and approaches the Poisson curve for large ones, for both models and both phases, consistently with lower edge modes becoming more and more localised for increasing sizes. In figure \ref{fig:pdf_r_fixed_relative_rank} instead we show the same plots but fixing the relative rank $\kappa=k(N)/(m-1)N=10^{-4}/(m-1)$ for growing sizes: with this choice, we can observe the statistics of $r$ for a fixed position in the spectrum, so we expect curves related to different sizes to collapse into a single one.
In the XY case we observe no significant difference between the paramagnetic and the spin glass phase: for the value of relative rank $\kappa=10^{-4}$, data are fairly consistent with the Poisson distribution in both phases.
On the contrary, for relative rank $\kappa=0.5\times 10^{-4}$ in the Heisenberg case the two phases seem to have different statistical properties: in the paramagnetic phase, the PDF of $r$ is again consistent with the Poisson distribution, but in the spin glass phase this is not the case, rather the PDF of $r$ seems to be intermediate between the Poisson and the Wigner surmises. This observation hints the presence of multifractal delocalised modes in the bulk of the spectrum, close to the lower edge. To further explore this possibility, we show in figure \ref{fig:r_avg} the sample average value of $r$ from our data as a function of the relative rank $k/(m-1)N$, where again for each size sample averages are performed over adjacent windows of ten eigenvalues. The plots in the left (XY) and in the right (Heisenberg) confirm the previous observations: in the Heisenberg model, the average value of $r$ in the spin glass phase is significantly larger than that in the paramagnetic phase for a wide range of low relative ranks; conversely, in the XY model the average value of $r$ differs between the two phases only for the largest ranks, being robustly Poissonian for the smallest ones.


\newpage

\section{Properties of Random Regular Graphs}
\label{sec:properties_random_regular}

Here we discuss some properties of random regular graphs. This appendix does not contain new results: we collect known facts about RRGs, for readers convenience.

\subsection{The average number of nodes at fixed distance}

Let us a consider a RRG with $N$ nodes and connectivity $c$. In a finite size random graph, the graph is locally tree-like, with an exact number of neighbors at topological distance $d$
\begin{equation}
\label{eq:growth_regular_tree}
    \mathcal{N}_{tree}(d)\,=\,c(c-1)^{d-1}(1-\delta_{d,0})+\delta_{d,0}
\end{equation}
up to a certain scale $O(\log_{c-1} N)$. With the appearance of loops, the number of neighbors at fixed distance departs from \eqref{eq:growth_regular_tree}, peaking approximately at $L\sim\log(N)/\log(c-1)$ and decreasing very fast for greater distances.

\begin{figure}[h]
    \centering
    \includegraphics[width=0.7\textwidth]{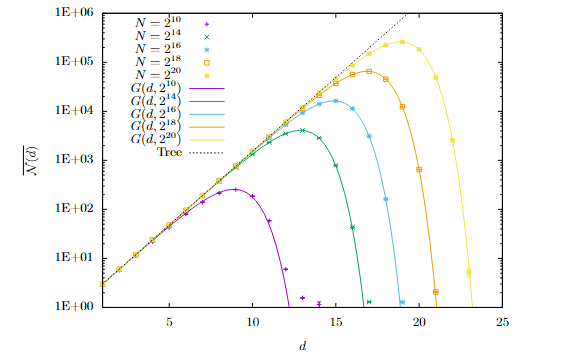}
    \caption{The average number of nodes at distance $d$ from a random node. Points represent numerical measures, whereas continuous curves come from the analytical prediction of eq. \eqref{eq:growth_RRG}, which in the legend of the plot is called $G(d, N)$. The agreement between numerical data and eq. \eqref{eq:growth_RRG} is extremely good. Significant deviations at large $d$ are present only for the smallest size shown.}
    \label{fig:n_RRG}
\end{figure}

For generic graphs, the average number of nodes at a given distance
\begin{equation}
    \mathcal{N}(d)=\overline{\frac{1}{N}\sum_{i\in \mathcal{V}}\sum_{j\in \mathcal{V}} \mathbf{1}(d(i, j)=d)}
\end{equation}
is not known through an explicit formula. In the case of random regular graphs (RRG), ~$\mathcal{N}(d)$
can be estimated exactly for large $N$. We can write this quantity as
\begin{equation}
    \mathcal{N}(d)\,=\,N\;p(d)
\end{equation}
where $p(d)$ is the probability that the distance between two random nodes is $d$: this quantity is usually called in literature as Distribution of Shortest Path Lengths (DSPL).
The tail DSPL $P(x>d)$ of RRG was originally derived in \cite{vanDerHofstad2005} and recently rederived in \cite{tishby2022mean} through cavity method: it is a discrete Gompertz distribution and reads
\begin{equation}
    P(x>d)=\exp\left\{-\frac{c}{(c-2)N}[(c-1)^{x-1}-1]\right\}
\end{equation}
Since $p(d)=P(x>d-1)-P(x>d)$,  the average number of nodes at distance $d$ of RRGs is
\begin{equation}
\label{eq:growth_RRG}
    \mathcal{N}(d)\,=\,Ne^{\eta}\left\{e^{-\eta(c-1)^{d-1}}-e^{-\eta(c-1)^{d}}\right\}
\end{equation}
where we set $\eta=\frac{c}{N(c-2)}$. In figure \ref{fig:n_RRG} we compare numerical measures of the average number of neighbors at distance $d$ in a RRG with formula \eqref{eq:growth_RRG}. Our data agree very well with \eqref{eq:growth_RRG}, having finite size effects at large distances for the smallest sizes.
We can estimate the scale $d_*$ such that for $d\ll d_*$ one has $\mathcal{N}(d)\approx \mathcal{N}_{tree}(d)$. It is easy to show that at first order in $\epsilon=\eta (c-1)^d\ll 1$ one has that for
\begin{equation}
   d\ll \frac{\log(N)}{\log(c-1)}-\frac{\log\left(\frac{c}{c-2}\right)}{\log(c-1)}=d_*
\end{equation}
eq. \eqref{eq:growth_RRG} reduces to eq. \eqref{eq:growth_regular_tree}. Let us compare $d_*$ with the distance $d_M$ at which $\mathcal{N}(d)$ defined by \eqref{eq:growth_RRG} has its absolute maximum: by imposing the null derivative of \eqref{eq:growth_RRG}, one finds the solution
\begin{equation}
    d_M=d_*+1+\frac{\log(\log(c-1))}{\log(c-1)}-\frac{\log(c-2)}{\log(c-1)}
\end{equation}
The maximum on the discrete distribution $d_m$ is the closest integer to $d_M$:
\begin{equation}
    d_m\,=\,[d_M]\theta\left(\frac{1}{2}-d_M+[d_M]\right)+([d_M]+1)\theta\left(d_M-[d_M]-\frac{1}{2}\right).
\end{equation}

\section{The role of less relevant eigenvector maxima}
\label{sec:less_relevant_eigenvector_maxima}
\begin{figure}[h]
    \centering
    \includegraphics[width=0.6\textwidth]{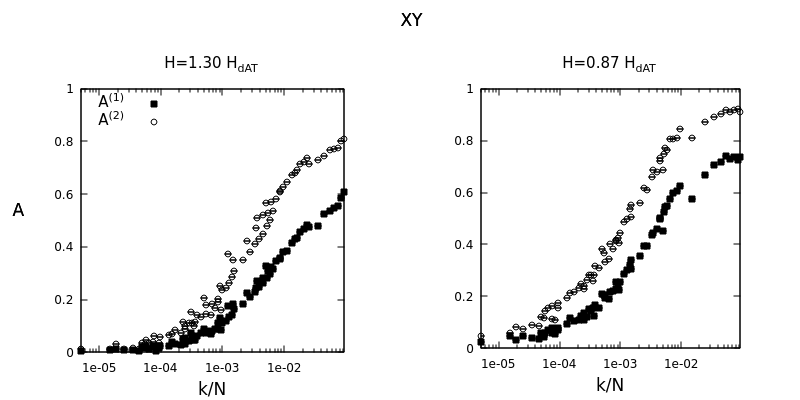}
    \includegraphics[width=0.6\textwidth]{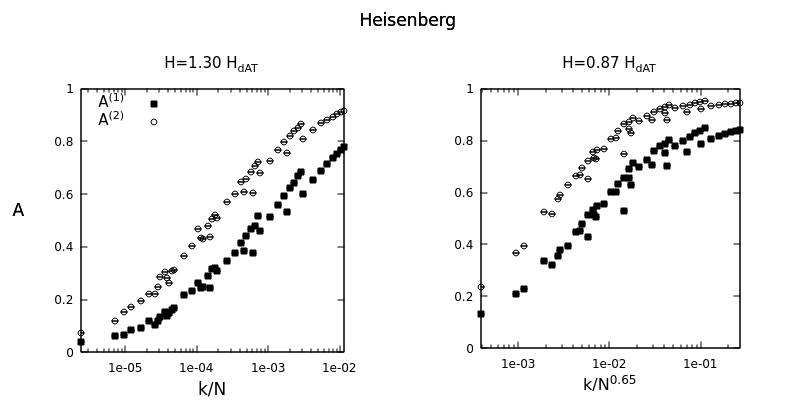}
    \caption{\textbf{Top}: the improved area for $H=1.30 \Hdat$ and $H=0.87 \Hdat$ in the case of the XY model, compared with the one defined by eq. \eqref{eq:Pgreater}.
    \textbf{Bottom}: the improved area for $H=1.30 \Hdat$ and $H=0.87 \Hdat$ in the case of the Heisenberg model, compared with the one defined by eq. \eqref{eq:Pgreater}.}
    \label{fig:Pgreater_improved}
\end{figure}

Here we refine the criterion used in subsection \ref{subsec:nongeom_sm} for the classification of CDLEM. We consider the role of the third maximum: we redefine the probability of observing a CDLEM with rank $k$ as
\begin{equation}
\label{eq:Pgreater_improved}
    \mathcal{A}^{(k;2)}=\mathbb{P}(d_{12}\geq [0.7 L])+\mathbb{P}(d_{13}\geq [0.7 L]\;\lvert\;d_{12}\leq [0.7 L])
\end{equation}
where
\begin{equation}
\label{eq:Pgreater}
  \mathcal{A}^{(k;1)}=\mathbb{P}(d_{12}\geq [0.7 L])  
\end{equation}
is the criterion used in subsection \ref{subsec:nongeom_sm} to identify CDLEM.
In figure \ref{fig:Pgreater_improved} we show the area defined by \eqref{eq:Pgreater}, $\mathcal{A}^{(k)}\equiv \mathcal{A}^{(k;1)}$, versus $k/N$, for $H=1.30 \Hdat$ and $H=0.87 \Hdat$. We show the curves related to the area defined by eq. \eqref{eq:Pgreater_improved} opaque. The qualitative behavior of the improved probability $\mathcal{A}^{(k, 2)}$ with respect to $k/N$ is the same. So, considering third maxima in the definition of CDLEM does not change our conclusion on the localisation properties of the two models.

\end{document}